\documentclass[aps,pra,
twocolumn,showpacs,psfig]{revtex4}

\newcommand{\eps}{\varepsilon}

\newcommand{\la}{\lambda}

\usepackage{graphicx}
\usepackage{amsmath}
\usepackage{amssymb}

\begin{document}
%\twocolumn[\hsize\textwidth\columnwidth\hsize
%\csname@twocolumnfalse%
%\endcsname
\draft
\title {On creation and evolution of dark solitons in Bose-Einstein
condensates}
\author{V.A. Brazhnyi$^{1}$}
\email{brazhnyi@cii.fc.ul.pt}
\author{A.M. Kamchatnov$^{1,2}$}
\email{kamch@isan.troitsk.ru}
\author{V.V. Konotop$^{1}$}
\email{konotop@cii.fc.ul.pt}
%\address{
\affiliation{$^1$Centro de F\'{\i}sica da Mat\'eria
Condensada, Universidade de Lisboa,
    Complexo Interdisciplinar, Av. Prof. Gama Pinto 2, Lisboa 1649-003,
         Portugal \\
         $^2$Institute of Spectroscopy, Russian Academy of Sciences,
Troitsk 142190, Moscow Region, Russia}
%\date{\today}


\begin{abstract}
Generation of dark solitons from large initial excitations and
their evolution in a quasi-one-dimensional Bose-Einstein
condensate trapped by a harmonic potential is studied analytically
and numerically. In the case of a single deep soliton main
characteristics of its motion such as a frequency and amplitude of
oscillations are calculated by means of the perturbation theory
which in the leading order results in a Newtonian dynamics,
corrections to which are computed as well. It is shown that
long-time dynamics of a dark soliton in a generic situation
deviates substantially from outcomes of the naive application of
the Ehrenfest theorem. We also consider three different techniques
of controllable creation of multi-soliton structures (soliton
trains) from large initial excitations and calculate their initial
parameters (depths and velocities) with the use of a generalized
Bohr-Sommerfeld quantization rule. Multi-soliton effects are discussed.
\end{abstract}

\maketitle

\section{Introduction}

Experimental observations of dark solitons in a cigar-shaped
Bose-Einstein condensate (BEC) of sodium \cite{andrews} and of
$^{87}$Rb \cite{Burger1} atoms and in a near spherical BEC of
$^{23}$Na atoms \cite{Densch} have stimulated intensive
theoretical studies devoted to generation and evolution of such
excitations in BEC's. At low enough temperature the condensate
evolution is described sufficiently well by the Gross-Pitaevskii
(GP) equation \cite{Pit}, which is originally three-dimensional
(3D) but in  some cases is reducible to a 1D nonlinear
Schr\"odinger (NLS) equation with an external potential. As such
cases we can mentioned a BEC with a ``pancake" geometry (see e.g.
\cite{pancake}) and a low-density BEC in a ``cigar-shape" trap
(see e.g. \cite{KS1}). If the respective size of the condensate is
much bigger than the characteristic dimension of an excitation in
it, then it is commonly believed that the conventional (i.e.
without potential) NLS equation can be used for understanding the
excitation evolution during initial intervals of time. Such
approximation was used for description of dark soliton dynamics in
quasi-1D BEC with a positive scattering length (see e.g.
\cite{solbezBG}). If, however, a size of an excitation is
comparable with the size of the condensate or if one is interested
in long-time dynamics, then influence of a trap potential must be
taken into account. This is especially important in the case of
dark solitons. One of the physical reasons for that is that in the
presence of a trap potential the BEC density becomes a function of
a space coordinate and/or time. From the mathematical point of
view it means that the trap potential changes boundary conditions
for the macroscopic wave function of the condensate at the
infinity. (Note that this does not happen in the case of bright
solitons in a BEC with negative scattering length, which in a
quasi-1D case were recently observed experimentally
\cite{Nature}).

In the homogeneous NLS equation, which is
exactly integrable~\cite{Zakh}, the long-time evolution of an excitation
can be reduced to the motion of a set of solitons and to propagation of
linear waves. Solitons in that case are well defined objects and their
parameters can be found by means of the inverse scattering transform method
\cite{Zakh}. In particular, constant velocities of solitons (as well as other
soliton parameters) can be calculated from the associated linear spectral
problem using the initial distribution of the macroscopic wave function. In the
presence of a trap this method cannot be applied without some reservations.
Moreover, a dark soliton against a nonuniform background is not  a well
defined object anymore. Therefore separation of the excitation evolution into
``soliton part" and ``background part" is somewhat conditional.
Even if such separation is meaningful at the initial moment of time,
during propagation soliton parameters become functions of time.
In particular,  a change of a soliton shape can be so dramatic that
the distinction between the soliton and the background may loose its sense
and must be reconsidered. Then one meets a problem of long-time description of
the soliton evolution. This problem becomes even more complicated in the case
when an initial excitation results in formation of many solitons (soliton trains).
Thus, one of the  aims of the present paper is to consider evolution of excitations
which initially can be classified as one- and  multi-soliton ones in 1D BEC confined
by a harmonic potential.

Several aspects of this problem have already been addressed in literature.
In particular, there have been reported several evidences that the dynamics of a
dark soliton or a few solitons in a BEC trapped in a harmonic potential is
oscillatory~\cite{Reinhard,Scott,Busch,Huang}.
In one of the first   publications~\cite{Reinhard} it has been suggested that
for description of the dark soliton evolution one can employ the Ehrenfest
theorem which allows one to define a frequency of oscillations, and on this
basis an empiric formula $m\ddot x =-\partial U/\partial x$,
where $U(x)$ is a trap potential, for Newtonian dynamics of a dark soliton
has been proposed (although a definition for the coordinate of a dark soliton
$x$ has not been given). A similar equation for a dark soliton motion against
a background was derived in \cite{Busch} by means of the multi-scale analysis.
In numerical simulations provided in \cite{Huang} authors observed an oscillatory
behavior of small amplitude dark solitons.
That paper however has left open a question about the frequency of the oscillations
and their amplitude. Also, although an important question about a choice of a
particular form of the background has been posed (since a ``wrong'' choice results
in rapidly growing oscillations of the background), it has been solved by numerical
means on the basis of elimination of the mentioned oscillations of the background,
leaving analytical fundamentals for such a choice open, as well. Moreover in
Ref. \cite{Huang} it has been found that the amplitude of a small-depth dark soliton
increases as the soliton approaches a turning point, which does not corroborate
with the previous findings \cite{Reinhard,Scott,Busch}
where it was reported that the shape of a dark soliton is preserved during
propagation against a background. Finally, it is worthwhile to mention that
in Refs.~\cite{Reinhard,Scott,Busch,Huang}  one can find no information about
long-time behavior of a dark soliton in a trap potential.

Another related issue of the theory is generation of dark solitons.
%As in the case of the propagation problem, existence of a trap  makes the
%problem very different from its counterpart in the NLS-soliton theory.
The discussion in the literature has been mainly concerned with such methods
as laser induced Raman transition between two internal states of the
condensate~\cite{Dum}, ``phase imprinting" \cite{Burger1,Biao}, modulational
instability \cite{KS1},  and the use collision of two initially separated
condensates \cite{Reinhard,Scott}.

In Ref. \cite{Reinhard}  it was supposed that generation
of the dark solitons with different initial velocities can be achieved
in a collision of two pieces of initially displaced condensates.
By numerical calculations it was shown that in the presence of a trap all solitons
have essentially the same classical period of oscillations given by harmonic
classical dynamics.
The same mechanism of generation of the multi-soliton structures was considered in
Ref.~\cite{Scott} (although a number of the generated solitons and their initial parameters were not calculated).
In Ref. \cite{Dum} a scheme to create dark solitons in a BEC with the use
of coherent Raman process which
couples internal atomic levels of the condensate with a laser was proposed.
In a recent work on the controlled generation of dark solitons by phase
imprinting method \cite{Biao} the authors provided a simple theoretical description
of the creation of the dark solitons in the experiment  \cite{Burger1}.
Considering a very short time scale, they neglected the trap potential,
and in this case for a given initial phase step, a number of solitons and their initial
velocities were calculated
analytically by mapping Zakharov-Shabat eigenvalue problem into the pendulum problem,
which is mathematically much simpler. %That approach however cannot be applied for description of the
%long-time evolution when a trap potential becomes a crucial factor.

Thus, the second aim of the present paper is the study of generation of multi-soliton
structures from an arbitrary initial excitation
in a nonuniform condensate confined by a harmonic potential as well as their evolution.

The organization of the paper is as follows. In Section II the dynamics of a single dark
soliton in a harmonic trap potential is considered in detail. First of all we argue
the choice of the analytical form of the background which is necessary for a stable
long-time dynamics of a soliton making it nearly integrable. We show also that
one can define two characteristic coordinates associated with the dark soliton:
a position of the center of mass (the mass being negative) and a position of a local
minimum of the intensity. The both values being the same in the case of a
homogeneous background display different behavior subject to the effect of the potential.
The first one is governed by the Ehrenfest theorem while another one can be described on the basis of the perturbation
theory of solitons~\cite{KV94}. Theoretical predictions are compared with the numerical
simulations. Non-adiabatic effects of the soliton dynamics and the behavior of its phase
are discussed.  In Section III we describe generation of soliton trains
in a trapped BEC from initial excitations of different types and study their evolution.
In the case of perturbation of the condensate density by a large
and smooth initial pulse we find initial parameters of created
solitons with the use of generalized Bohr-Sommerfeld quantization rule~\cite{Kamch1}.
Using the results of Section II we predict locations of turning points which are
well confirmed by direct numerical simulations. We also discuss behavior
of dark solitons generated by the phase imprinting method and creation of solitons
during collisions of two condensates in the presence of the harmonic trap potential.
The outcomes of the theory are summarized in Conclusion. For the sake of convenience a
summary of some technical results of the perturbation theory for dark solitons is given
in the Appendix.

\section{Motion of a single dark soliton in a parabolic trap}

\subsection{Statement of the problem}

As it is customary, we start with the GP equation for the order parameter
$\psi\equiv \psi({\bf r},t)$:
\begin{equation}
\label{GPE}
i\hbar \frac{\partial\psi}{\partial t}=-\frac{\hbar^2}
{2m}\Delta\psi+V_{trap}({\bf r})\psi+g_0|\psi|^2\psi\, ,
\end{equation}
where we use the standard notations: $g_0=4\pi\hbar^2a_s/m$, $a_s$ being
the $s$-wave scattering length, which is considered positive, and $m$ being the atomic mass;
$ V_{trap}({\bf r})$ is a trap potential.

A self-consistent reduction of Eq. (\ref{GPE}) to a 1D NLS equation can be made
in various situations (see e.g. \cite{pancake,KS1}).

(i) In the case of a pancake BEC we suppose that in the transverse
direction (i.e. in the direction orthogonal to the $x$-axis) the size of the condensate
is large enough to be considered infinite in the first approximation. This leads to
the trap potential of the form $V_{trap}=\frac m2 \omega_0^2x^2$ where $\omega_0$
is the harmonic oscillator frequency. Then, in order to rewrite the dynamical
equation in a dimensionless form we make a substitution
\begin{equation}
\label{subs}
\psi (\mathbf{r},t)=2^{\frac 14}(4\pi \nu a_sa_0^2)^{-\frac 12}
\exp\left(i{\bf k}_\bot {\bf r}_\bot-i\frac{\hbar k_\bot^2}{2m}t\right)\Psi(x,t),
\end{equation}
where ${\bf r}_\bot=(y,z)$ and  $a_0^2=\frac{\hbar}{m\omega_0}$, and make a change of independent variables
$x\mapsto \nu^{1/2} a_0 x/2^{1/4}$, $t\mapsto 2^{1/2}\nu t/\omega_0$,
 which results in the canonical form of the  NLS equation with a parabolic potential
\begin{equation}\label{NLS}
i\frac{\partial\Psi}{\partial t}+\frac{\partial^2\Psi}{\partial{x}^2}
-2 |\Psi |^2\Psi =\frac12\nu^2x^2\Psi
\end{equation}
(in what follows $\Psi$  is also referred to as a macroscopic wave function).
Notice that in contrast to the generally accepted renormalization, here we
introduced a parameter
$\nu$ which on the one hand characterizes a strength of the parabolic potential
in dimensionless equation (\ref{NLS}), and on the other hand is connected with the density
of particles in the condensate. To elucidate this connection, we note that
the condensate wave function $\Psi(x,t)$ is normalized according to
\begin{equation}
\label{normaliz}
\int_{-\infty}^{\infty} |\Psi(x,t)|^2dx=4\pi 2^{-1/4}\nu^{1/2} a_sa_0n_0,
\end{equation}
where $n_0$ stands for the ``transverse density'' of particle,
i.e. $n_0={\cal N}/S$, ${\cal N}$ is the total number of
particles and $S$ is the area of the transverse cross section of the condensate.
(Formally one has to consider the ``thermodynamical"
limit ${\cal N}\to\infty$, $S\to\infty$ at $n_0=$const.)

On the other hand let us consider an unperturbed condensate wave function $\Psi(x,t)=\exp(-i\mu t)\Psi(x)$ ($\mu$ is a renormalized chemical potential) which solves the stationary equation
\begin{eqnarray}\label{NLS-stat}
&&\frac{d^2\Psi}{d{x}^2}+\mu\Psi -2 |\Psi |^2\Psi =\frac12\nu^2x^2\Psi,
\\
\nonumber
&&\Psi(0)=1, \quad \Psi'(0)=0.
\end{eqnarray}
This is clear that $\Psi$ decays exponentially at $|x|\to\infty$, and thus one can expect that
\begin{eqnarray}
\label{TF0}
\int_{-\infty}^{\infty}|\Psi(x)|^2dx
\simeq \int_{-\sqrt{2\mu}/\nu}^{\sqrt{2\mu}/\nu}|\Psi_{TF}(x)|^2dx
=\frac{(2\mu)^{3/2}}{3\nu}
\end{eqnarray}
where
\begin{equation}
\label{TF}
\Psi_{TF}(x)=\frac 12\sqrt{2\mu-\nu^2x^2}
\end{equation}
is the condensate wave function in the Thomas-Fermi approximation
and $\mu\simeq 2$ for $\nu \ll 1$. More careful numerical study of Eq.~(\ref{NLS-stat})
gives corrections to (\ref{TF0})
\begin{equation}\label{b2}
\int_{-\infty}^{\infty} |\Psi(x)|^2dx=\frac{C(\nu)}{\nu},
\end{equation}
where $C(\nu)$ is a very slow function
of $\nu$ equal to $C(\nu)\simeq 2.67$ with accuracy better than
$1\%$ in the interval $0.1\leq\nu\leq0.3$. Comparison of Eqs.~(\ref{normaliz})
and (\ref{b2}) shows that
\begin{equation}\label{b3}
  \nu\simeq \left(\frac{2.67\cdot 2^{1/4}}{4\pi}\right)^{2/3}\frac{1}{(a_s a_0 n_0)^{2/3}}
  \simeq \frac{0.4}{(a_sa_0 n_0)^{2/3}}.
\end{equation}
This formula determines $\nu$ in terms of the physical parameters
of the system under consideration. As an example one can consider
${\cal N}=10^5$ atoms of $^{87}$Rb ($a_s\approx 5.8\,$nm) condensed in
a trap with $a_0\approx 1\mu$m (what corresponds to the frequency
$\omega_0\sim 7\cdot 10^2\,$Hz) and with transverse radius $\approx
14\mu$m. Then we get $\nu\approx 0.2$. As it is evident,
increasing of the density by a factor 10 results in decreasing of
the small parameter by approximately the same factor.

Thus, $\nu$ can be viewed as a natural small parameter of the problem and therefore
in what follows we shall consider the case  $\nu\ll 1$.

In order to construct the theory yet another condition is to be imposed:
the longitudinal dimension of the condensate $\sim \nu^{-1}$ must be sufficiently large
compared with a characteristic width $l\sim 1$ of a typical soliton solution.
Then it makes sense to speak about a soliton against an inhomogeneous
background. More specifically, we require
$
l/a_0\sim\sqrt{\nu}.
$
  This however is not a strong restriction since already at
$\nu\approx 0.1$ the relation $l\approx a_0/3$ (which corresponds to the
actual experimental settings) is verified.

(ii) Considering a cigar-shaped BEC one uses the multi-scale expansion
(see e.g. \cite{KS1} for details) with respect to the small parameter
$\epsilon=8\pi{\cal N}a_s a_\bot /a_0^2\ll 1$ (it can be viewed as smallness
of the energy of two-body interaction compared with the kinetic energy)
where $a_\bot=(\hbar/m\nu_\bot)^{1/2}$
is the linear oscillator length in the transverse direction and $\nu_\bot$
is the linear oscillator frequency in the transverse direction. The smallness of
the parameter $\epsilon$ provides the conditions when the consideration can be
restricted by the lowest mode in the three-dimensional parabolic trap: at
larger $\epsilon$ interactions of modes become essential and must be taken
into account what leads to a set of coupled nonlinear Schr\"{o}dinger equations.
In this case the time is measured in units $2/\nu_\bot$ and spatial coordinate along
the cigar axis is measured in units $a_\bot$. The parameter $\nu$ is now defined
as $\nu=a_\bot /a_0$ and can be easily made as small as necessary. Meantime, in
that case one has to verify the condition $\epsilon\ll 1$ necessary for applicability of
the theory developed below. Let us do that for the data analogous to ones reported
in \cite{andrews} imposing however a lower particle density, i.e.
considering a BEC of ${\cal N}=5\cdot 10^5$ sodium atoms ($a_s\approx 2.7$nm).
Then for $a_0\approx 450\,\mu$m and $a_\bot\approx 10\,\mu$m one estimates
$\nu\approx 0.1$. Thus in the case
at hand the limit of small effective parameter $\nu$ is also reachable for
available experimental settings.

In order to complete the statement of the problem we have to
specify the terminology ``long-time'' dynamics. We will do this
for a pancake geometry. A unity of time, $\Delta t=1$, in the
dimensionless variables corresponds to $\omega_0/\nu$ real
seconds. This means, for example, that considering a BEC of
$^{87}$Rb atoms in a trap with the longitudinal size of order of
$a_0\sim 1\mu$m and assuming that $l/a_0\sim 0.3$, a unity of the
dimensionless time will correspond to 0.18 ms. A typical life-time
of a BEC reachable in today experiments can be estimated as at
least 50 ms, which means that it is meaningful (in a BEC context)
to investigate the dynamics of the excitations for times at least
up to 300 dimensionless units. As one expects $\nu$ to be of order
of the frequency of soliton oscillations in a trap, the above
estimate corresponds to approximately 20 periods of oscillations.
Thus, long-time dynamics in the present paper is understood as a
dynamics displaying more than 10 oscillations.

\subsection{Choice of the background and near-integrable dynamics}

When $\nu=0$, Eq. (\ref{NLS}) reduces to the conventional NLS equation which
admits a stable constant-background solution $|\Psi(x,t)|^2=\rho_0$(=const),
against which one can construct
a dark soliton solution \cite{Zakh}
\begin{equation}\label{adiab_sol}
\Psi_s(x,t)\equiv\sqrt{\rho_0}\,
\frac{1+e^{i\vartheta}\exp\{\eta_0 [x-X(t)]\}}{1+\exp\{\eta_0 [x-X(t)]\}}.
\end{equation}
Here $X(t)=Vt+X_0$, $V = -2\sqrt{\rho_0} \cos(\vartheta/2)$ and
$\eta_0=2\sqrt{\rho_0}\,\sin(\vartheta/2)$
are the soliton coordinate, velocity, and width, respectively. $X_0$
is the coordinate at initial moment of time.
Due to the symmetry of the problem, the parameter $\vartheta$ which
characterizes the phase difference of $\Psi_s(x,t)$ at $\pm\infty$ can be restricted to
the interval $0\leq\vartheta\leq\pi$. For $\vartheta=\pi$ the BEC density
vanishes at the soliton center and in this case a soliton is often called
``black'' while for other $\vartheta$ a soliton is referred to as ``grey''. The limit
$\vartheta\ll 1$ corresponds to small amplitude solitons.

When the trap potential is switched on, the function $\Psi_s(x,t)$
given by (\ref{adiab_sol}) is not a solution of (\ref{NLS}) anymore.
However, one could expect that if a region of soliton localization is
much less than the size of the background, i.e. when  $\nu\ll\eta_0$, and if in
the vicinity of the potential center the initial
condition for Eq.~(\ref{NLS}) is chosen to be close enough to the dark
soliton one, then a solution of Eq.~(\ref{NLS}) can be searched in a
form of a ``dark soliton'' $\Phi(x,t)$ against an inhomogeneous background
$F(x)$, i.e. in the form
\begin{equation}
\label{factor}
  \Psi(x,t)=F(x)\cdot\Phi(x,t) ,
\end{equation}
where $\Phi(x,0)=\Psi_s(x,0)$, $\Psi_s(x,t)$ being given by (\ref{adiab_sol}).

Ansatz (\ref{factor}) is meaningful if the dynamics of $\Phi(x,t)$ is close,
in some sense, to the dynamics of $\Psi_s(x,t)$. Since $\Psi_s(x,t)$ is governed
by the unperturbed (i.e. with $\nu=0$) NLS equation, to satisfy the last
requirement it is natural to choose $F(x)$ such that the resulting equation
for $\Phi(x,t)$ would be close to the NLS equation. This can be achieved by
requiring $F(x)$ to be an eigenfunction of the nonlinear spectral problem
[c.f. (\ref{NLS-stat})]
\begin{eqnarray}
\label{eq8}
F_{xx} + \left(\omega_b - \frac{(\nu x)^2}{2}\right)F - 2\rho_0 F^3=0, \\
\label{eq8a}
\lim_{x\to\pm\infty} F(x)=0,
\end{eqnarray}
which satisfies the following
normalization conditions
\begin{eqnarray}
\label{eq8b}
F(0)=1, \qquad F_x(0)=0.
\end{eqnarray}
In (\ref{eq8}) $\omega_b$ is an eigenvalue.

Indeed, substituting ansatz (\ref{factor}) in Eq.~(\ref{NLS}), one obtains
\begin{equation}\label{pert_NSE}
i \Phi_t+\Phi_{xx}-2 (|\Phi |^2-\rho_0)\Phi = R[\Phi, F],
\end{equation}
where
\begin{equation}\label{R}
R[\Phi, F]\equiv -2\left(\ln F\right)_x \Phi_x+2\Phi(|\Phi|^2-\rho_0)(|F|^2-1).
\end{equation}
Next, one can estimate the order of magnitude of $R[\Phi, F]$
considering $x\nu$ small enough:
\begin{eqnarray}
\label{gen_est}
\displaystyle{
\begin{array}{l}
\left(\ln F\right)_x\sim \nu^2x,\quad
\Phi_x\sim \eta,
\\ (|\Phi|^2-\rho_0)(|F|^2-1)\sim\nu^2\eta^2x^2,
\end{array}
}
\end{eqnarray}
where it is taken into account that in a vicinity of the center of
the potential, i.e. at $x\nu\ll 1$, $F(x)$
can be expanded in the Taylor series
\begin{equation}\label{expans_F}
F(x)=1+\frac{2\rho_0-\omega_b}{2}x^2+O(x^4)\,.
\end{equation}
Substitution of (\ref{expans_F}) into (\ref{eq8}) with subsequent expansion in
powers of ``small" $x$ (i.e. $x=o(\nu^{-1})$) and
small $\nu$ yields the estimate for the eigenvalue
\begin{equation}\label{rez}
\omega_b=2\rho_0+\frac{\nu^2}{4\rho_0}+O(\nu^4).
\end{equation}

As it follows from the definition of the soliton width, we have
$\eta=O(1)$ whenever $\rho_0=O(1)$.
This means that  $\nu\ll 1$ is indeed a small parameter of the problem in
the sense that $R[\Phi, F]=O(\nu^2)$.
Then the evolution of the function $\Phi(x,t)$ will be described by nearly integrable
equation (\ref{pert_NSE})
and in this sense will be sufficiently close to the evolution of $\Psi_s(x,t)$
given by (\ref{adiab_sol})
at least for period of time defined by $t\ll \nu^{-4}$.
We will refer to the respective dynamics as {\em nonlinear}. If, however, $x\nu\gtrsim 1$
then the first term in (\ref{R}) becomes exponentially small (since as it follows
from (\ref{eq8}) decay of $F(x)$ at $x\to\pm\infty$ is exponentially fast) and the second term is approximately canceled with the nonlinear term in (\ref{pert_NSE}).
In that case the dynamics is governed by the effectively linear equation.

\subsection{Nonlinear dynamics of a dark soliton.}
\label{nonlin_dynam}

Suppose that $X(t)=o(\nu^{-1})$. In this case estimates (\ref{gen_est})
hold for the whole spatial region and
one can  simplify the expressions for the background with the use  of
(\ref{expans_F}) and (\ref{rez}):
\begin{equation}
\label{f_expan}
F=F_0(x)+O(\nu^4x^4),\qquad F_0(x)\equiv1-\frac{\nu^2}{8\rho_0}x^2
\end{equation}
and rewrite $R[\Phi,F]$ in the form
\begin{eqnarray}
\label{RP1}
R[\Phi,F]\approx R[\Phi,F_0]
\equiv \frac{\nu^2x}{2\rho_0}\left[\Phi_x - x\Phi(|\Phi|^2-\rho_0)\right]\,.
\end{eqnarray}

Now Eq. (\ref{pert_NSE}) can be treated by means of the perturbation theory for
dark solitons~\cite{KV94}. In this way, however, one meets a problem: the term
proportional to $x\Phi_x$ in the right hand side of (\ref{RP1})
belongs to the class of perturbations which effect on the soliton dynamics
cannot be described by the adiabatic approximation only.
To avoid this difficulty, we make an ansatz
\begin{equation}
\label{ans}
\Phi=\phi-i\frac{\nu^2}{2\rho_0}f(t)\frac{\partial\phi}{\partial x}\, ,
\end{equation}
where
\begin{equation}
\label{f}
f(t)=\int^t_0X(t')dt'+f_0,
\end{equation}
and the constant $f_0$ is to be determined later.

Necessity of the renormalization (\ref{ans}) can be justified by full
 invoking the perturbation theory (see Ref. \cite{KV94} for details).
 In order to give here some simple indication on the origin of the phenomenon,
 we notice that secular terms appearing due to perturbation of a nonlinear
equation are eliminated in the so-called adiabatic approximation which is
obtained from the exact soliton solution by allowing parameters to be
dependent on time. If in our case one substitutes $e^{-i\vartheta/2}\Psi_s(x,t)$,
where $\Psi_s(x,t)$ is given by (\ref{adiab_sol}), into the left hand side of
(\ref{NLS}) and computes the imaginary part, one ensures that it is an even
function of the spatial coordinate. Thus if Im$\,e^{-i\vartheta/2}R[\Phi,F]$
has an odd (with respect to $x$) component, a secular terms originated by such
a term cannot be eliminated by any modification of the soliton parameters. This
requires introducing additional phase factor $\varphi(x,t)$ [see (\ref{expan_phi}) below].
The peculiarity of such a phase shift, however, is that the imaginary part of
the respective term (after multiplying by $\exp(-i\vartheta/2)$) is proportional
to $\tanh \eta(x-X(t))$ and thus is zero at $x=X(t)$. Thus
if Im$\,e^{-i\vartheta/2}R[\Phi,F]$ is not zero at $x=X(t)$, it cannot be
eliminated by any modification of the adiabatic theory. The ansatz (\ref{ans})
allows one to count the mentioned ``dangerous'' term explicitly. Taking into account the explicit form of that term, namely the fact that it is related to translational invariance of the system and is localized about the dark soliton kernel, it can be interpreted as an {\em internal mode} of a dark soliton excited by spatially varying background.

Now the dynamical equation for $\phi$ with the accuracy $O(\nu^4)$ reads
\begin{eqnarray}
\label{phi_new}
i\phi_t+\phi_{xx}-2(|\phi|^2-\rho_0)\phi=\nu^2\widetilde{R},
\end{eqnarray}
where
\begin{eqnarray}
\widetilde{R}&=&\frac{1}{2\rho_0}\left[(x-X(t))\phi_x-
x^2(|\phi|^2-\rho_0)\phi \right.\nonumber\\
\nonumber
&+& 4i f(t)\phi^2\bar{\phi}_x+\frac{\nu^2f^2(t)}{\rho_0}
\left(2\phi |\phi_x|^2 - \bar\phi (\phi_x)^2 \right) \\
&-&\left. i\frac{\nu^4 f^3(t)}{2\rho_0^2}\, \phi_x |\phi_x|^2\right]\label{R1}
\end{eqnarray}
where the terms proportional to the second and third order of $f(t)$
in some cases may be not small (see below).

The perturbation theory can be applied to (\ref{phi_new}), (\ref{R1}). To this
end we pass to new independent variables $(x,t)\to(\Theta,t)$ where
\begin{equation}
\label{Theta_X}
\Theta = \eta(t)(x-X(t))\quad \mbox{and} \quad
X(t)= Vt+x_0(t)
\end{equation}
and $\eta(t)$ and $x_0(t)$ are allowed to be dependent on time with the
initial conditions $\eta(0)=\eta_0$ and $x_0(0)=X_0$, and
look for $\phi$ in the form
\begin{equation}
\label{expan_phi}
\phi(x,t)=e^{i\nu^2\varphi(x,t)}\left(\phi_{ad}+\nu^2\phi_1+\cdots\right)\, ,
\end{equation}
where
\begin{eqnarray}\label{a_Phi}
\phi_{ad}=\sqrt{\rho_0}e^{i\frac{\vartheta}{2}}
\left(\cos\frac{\vartheta}{2}+i\sin \frac{\vartheta}{2}\tanh \frac{\Theta}{2}\right)
\end{eqnarray}
is the adiabatic approximation.
As it has been mentioned above, $\eta(t)$ and $x_0(t)$ are (slow) functions
of time, describing variations of the width/depth and velocity of
the soliton, respectively.  Quantity $X(t)$, which gives a position of the
minimum of the soliton intensity must be interpreted as a {\em coordinate of the
soliton center}. By introducing the rapidly varying phase $\varphi(x,t)$ one can satisfy the equation of balance of the momentum~\cite{KV94}
\begin{equation}
\label{momentum}
P=\frac{1}{2i}\int_{-\infty}^{\infty}(\phi_\Theta\bar{\phi}-\phi\bar{\phi}_\Theta)d\Theta,
\end{equation}
which reads
\begin{equation}
\label{mom_bal}
\frac{dP}{dt}=-\int_{-\infty}^{\infty}(\widetilde{R}
\bar{\phi}_\Theta+\bar{\widetilde{R}}\phi_\Theta)d\Theta
\end{equation}
(all functions in the integrands are considered to be dependent on the
variables $(\Theta,t)$).

In order to find dependence of $\eta(t)$ on time, one can use the conservation law
\begin{equation}\label{conserv_low}
\frac{d\tilde{N}}{dt}=2\int_{-\infty}^{\infty} dx \,{\rm Im}(\bar{\phi}\widetilde{R})\, ,
\end{equation}
where
\begin{equation}\label{N}
\tilde{N}=\int_{-\infty}^{\infty} dx\, (\rho_0-|\phi|^2)\, .
\end{equation}
Direct calculations allow one to ensure that (\ref{conserv_low}), (\ref{N})
give $d\eta/dt=0$, and thus the amplitude of the soliton is conserved:
$\eta\equiv\eta_0$. This results gives an explanation of the numerical
findings reported in \cite{Reinhard,Scott,Busch}. It is important, however,
that it has been obtained in the limit $\eta\gg\nu$ and thus cannot be
compared with the outcomes of \cite{Huang}.

Now, computing $\widetilde{R}$ and substituting the result in (\ref{momentum})
and (\ref{mom_bal}),
one obtains subject to the boundary conditions $\lim_{|x|\to\infty}\varphi(x,t)=0$:
\begin{eqnarray}
\label{phi}
\varphi=-\frac{\nu^2\eta_0}{4\rho_0}\frac{\Theta}{\cosh^2(\Theta/2)}
\int_0^t  X(t^\prime)\,dt^\prime\, .
\end{eqnarray}

Let us  consider a soliton trajectory.
After direct substitution of Eq.(\ref{R1}) into Eqs.(\ref{a4}), (\ref{a5}) and
then substitution of the obtained result into (\ref{a1}) one can get (see Appendix A)
\begin{eqnarray}
\frac {dx_0}{dt}&=&
-\frac{\nu^2}{3}\left(2\sin^2\frac{\vartheta}{2}+1\right)\int_0^{t} X(t^\prime)\, d t^\prime
\nonumber\\
&-& \frac{\nu^2 V}{8\rho_0}X^2-\frac{4\nu^4 V\sin^2\frac{\vartheta}{2}}{9\rho_0}
\left(\int_0^{t} X(t^\prime)\, d t^\prime\right)^2
\nonumber\\
&+&
\frac{2\nu^6 \sin^4\frac{\vartheta}{2}}{15\rho_0}
\left(\int_0^{t} X(t^\prime)\, d t^\prime\right)^3.
\label{a6}
\end{eqnarray}
Requiring the initial soliton velocity to be $V$, i.e. $\frac{dx_0}{dt}|_{t=0}=0$ one obtains the constant introduced in (\ref{f}):
$
f_0=
\frac{(6+\pi^2)}{24}\frac{V}{\rho_0\eta_0^2}+O(\nu^2V^2).
$

Below we will argue that the theory is applicable for relatively small velocities,
in particular for $V\sim\nu$.
This means that we are to restrict the consideration to $\vartheta$ close to $\pi$.
Let us first consider
$|\vartheta-\pi|\sim\nu$.
This allows one  to verify, using Eq. (\ref{a6}),  that $X(t)$ in the leading
order with respect to $\nu$ (designated below as $X_0(t)$) satisfies the equation
of the harmonic oscillator
\begin{equation}
\label{Ehrenf2}
\frac{d^2 X_0}{ dt^2}+\nu^2 X_0(t)=0
\end{equation}
and thus
\begin{equation}
\label{sol}
  X_0(t)=\frac{V}{\nu}\sin(\nu t)
\end{equation}
(where the initial conditions $X_0(0)=0$, $\dot X_0(0)=V$ have been used).
Thus, the soliton center, defined as a local minimum of the intensity undergoes
oscillatory motion with the frequency which is equal to the frequency of the trap $\nu$.
This result also corroborates with previous investigations \cite{Reinhard,Busch,Huang}.
Notice that from Eq. (\ref{sol}) and the condition $|X(t)|\lesssim 1$ follows the
relation $V\sim\nu$ used above.

Let us now return to expansion (\ref{f_expan}) and observe that it is still valid if we
consider $X\sim \nu^{-1/2}$ (and $\nu^{1/2}\ll 1$). In this case the soliton velocity
is allowed
to be of order of $\nu^{1/2}$. Then differentiation of (\ref{a6}) with respect to
time yields (instead of (\ref{Ehrenf2}))
\begin{equation}
\label{corrections}
\frac{d^2 X}{ dt^2}+\tilde\nu^2 X(t)=\frac{\nu V^3}{\rho_0}\left(\frac{23}{72}\sin(2\nu t)
+\frac{1}{10}\sin(3\nu t)\right)
\end{equation}
where
\begin{equation}
\label{tilde_nu}
\tilde{\nu}^2=\nu^2\left(1+\frac{2}{9}\frac{V^2}{\rho_0}\right)
\end{equation}
is the renormalized frequency and in
the right hand side $X(t)$ has been substituted by $X_0(t)$ given by (\ref{sol}). Then one computes
\begin{eqnarray}
\label{corrections1}
X(t)&=&\left(1+\frac{541}{2160}\frac{V^2}{\rho_0}\right)
\frac{V}{\nu}\sin(\tilde\nu t)\nonumber\\
&-&\frac{V^3}{\rho_0\nu}\left(\frac{23}{216}\sin(2\nu t)+\frac{1}{80}\sin(3\nu t)\right).
\end{eqnarray}

The obtained result reveals three important features of the dynamics when
the soliton velocity
increases. First, the frequency of oscillations increases compared with the frequency of the
harmonic trap (in the case at hand by $\frac{1}{9}\frac{V^2}{\rho_0}$).
Second, the amplitude of oscillations also increases (by the value
$\frac{541}{2160}\frac{V^2}{\rho_0}$).
Third, there exists a small ($\sim\nu$) frequency mismatch between second harmonic and a double first harmonic and  between third harmonics and the triple first harmonic.
This leads to slow (compared with the period of soliton oscillations)
variation of the amplitude of the periodic motion.
Below we will observe all these phenomena in numerical simulations (see Fig.~\ref{figone}).

\subsection{On the Ehrenfest theorem for dark solitons}
\label{Ehrenfest}

As was indicated in \cite{Reinhard,Busch}, the frequency of the soliton oscillations
can be found with help of the Ehrenfest theorem which must be written for the
{\em ``center of gravity'' of the whole condensate}
\begin{equation}\label{mean}
\overline{x}(t)=\frac{1}{N}\int_{-\infty}^{\infty} x |\Psi(x,t)|^2 dx,
\end{equation}
where
\begin{equation}
\label{numb_part}
N=\int_{-\infty}^{\infty}  |\Psi(x,t)|^2 dx
\end{equation}
is a renormalized number of particles of the whole condensate ($N\sim{\cal N}$).
Then the coordinate of the dark soliton can be defined as
\begin{equation}\label{meanDS}
\overline{x}_s(t)=\frac{1}{N_s}\int_{-\infty}^{\infty} x [F(x)]^2(|\Phi(x,t)|^2-\rho_0) dx,
\end{equation}
where
\begin{equation}
\label{numb_part_DS}
N_s=\int_{-\infty}^{\infty}  (|\Phi(x,t)|^2-\rho_0)\, dx
\end{equation}
can be interpreted as a {\em negative mass} of a dark soliton.

One can easily show that $\overline{x}_s(t)$ coincides with $X_0(t)$, defined in
subsection~\ref{nonlin_dynam}, only in the leading order and at initial stages
of evolution, and thus cannot be treated as a coordinate of the soliton center
in a general case.
Indeed, it is straightforward to show that $\overline{x}_s(t)$
obeys the exact ``Newton law''
\begin{equation}\label{newton1}
\frac{d^2\overline{x}_s}{dt^2}=-\int_{-\infty}^{\infty}\frac{\partial V_{trap}}
{\partial x}|\Psi(x,t)|^2dx,
\end{equation}
which in the case at hand is reduced to the equation for the harmonic oscillator.
Then, taking into account the initial condition in the form \begin{equation}\label{ini}
\Psi(x,0)=F(x)\cdot\Phi_s(x,0),
\end{equation}
which corresponds to the form of the solution given by (\ref{factor}), one computes
\begin{equation}
\label{sol_x}
\overline{x}_s(t)=\frac{V}{\nu}\sin(\nu t),
\end{equation}
which is an exact formula (notice that Eq.~(\ref{sol}) is approximate).

Thus, we have obtained that in the lowest order of $\nu$ a dark soliton
with a small velocity $V\sim \nu$ undergoes oscillatory behavior
which emerges both from the perturbation theory for solitons and from
the Ehrenfest theorem. Such a behavior is confirmed by the direct numerical simulations
(see Fig.1c below, where $\nu=0.2$, and $V\approx 0.14$).
However, significant deviations from the Ehrenfest theorem are observed
with growth of the initial dark soliton velocity [c.f. (\ref{sol_x}) and
(\ref{corrections1})].

\subsection{Numerical simulations of the one-soliton dynamics.}

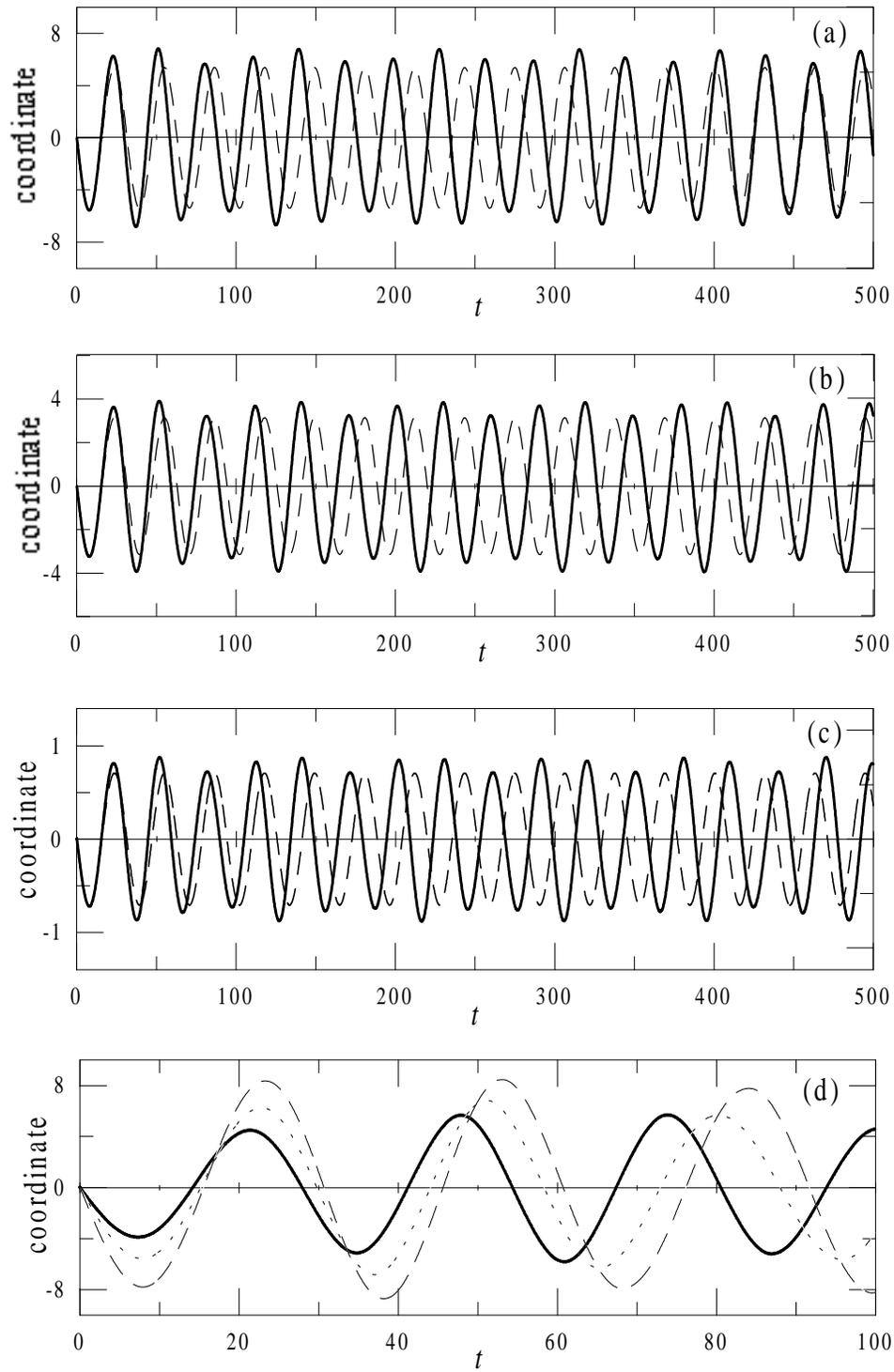
\begin{figure}[h]
%\centerline{\includegraphics[width=9cm,height=9cm,clip]{fig1.eps}}
%\centerline{\includegraphics[width=9cm,height=3cm,clip]{fig1d.eps}}
%\vspace{-1 true cm}
%\scalebox{.6}[0.6]{\includegraphics{fig1.eps}}
%\includegraphics{fig1.eps}
\vspace{0.3 true cm}
\caption{Dark soliton trajectories $X(t)$ (solid lines):
for $\rho_0=1$ and different initial parameters (a) $\vartheta_0=2$; (b) $\vartheta_0=2.5$; (c) $\vartheta_0=3$,
and the trajectories following from the Ehrenfest theorem $\bar{x}_s$
(dashed lines);
Figure (d) shows soliton trajectories  for $\vartheta_0=2$ and different amplitudes $\rho_0 =0.5;1;2$ (solid,  dotted and dashed line correspondingly).
The parameter of harmonic trap potential is  $\nu=0.2$.
}
\label{figone}
\end{figure}

To verify the above findings, we have done extensive numerical simulations of
the dark soliton dynamics
which is governed by the evolution equation Eq.~(\ref{NLS}) subject to the
initial conditions (\ref{ini}).
In Fig.~\ref{figone} we present a dark soliton trajectories $X(t)$, which
numerically were defined as the coordinate of the local minimum of the soliton,
from the long-time simulations. Although, strictly speaking, only the case depicted
in Fig.~\ref{figone}c corresponds to the perturbation theory developed above,
all the plots display several features qualitatively coinciding with the
theoretical predictions.

First, the choice of the form of the background
as an eigenfunction of the problem (\ref{eq8}) indeed allows one to follow the
dynamics up to hundreds of periods of oscillations, if initially soliton had large
enough amplitude, i.e. one still can identify  an entity called dark soliton.

Second, the dynamics of such defined $X(t)$ displays relatively large deviations
from that, which would follow from the
Ehrenfest theorem, when $\bar{x}_s$ is identified with the soliton coordinate.
Namely, in all the cases the frequency of the soliton oscillations is higher
than the frequency, predicted from the Ehrenfest theorem (see (\ref{sol_x})).
The observed frequency change is of order of a few percents
($\sim 3$\% in Fig.~\ref{figone}b) which agrees with prediction (\ref{tilde_nu}).

Third, one observes that the amplitude of
oscillations are larger then $V/\nu$ predicted by Ehrenfest theorem and
displays the periodic modulation.
All these differences increase with growth of the initial soliton velocity $V$.
This behavior is also in qualitative agreement with the result (\ref{corrections1})
obtained within the framework of the perturbation theory.

It is to be emphasized that all above effects become easily visible after several
periods of oscillations, while during the first period they closely follow the
``Ehrenfest trajectory'' given by (\ref{sol_x}) or (\ref{sol}). This coincidence
of the trajectories has been observed in Refs. \cite{Huang,Reinhard,Scott,Busch}
where only few cycles were counted.
Our numerical simulations of first  turning point also have shown that if an
initial soliton is deep enough (what corresponds to small values of $V$),
then the amplitude $a_{theor}=V/\nu$ found from
Eq.~(\ref{sol}) and amplitude $a_{num}$
calculated numerically as a local minimum of the intensity practically coincide
with each other, which is illustrated by Table~\ref{tabone}.

\begin{table}[h]
\caption{Amplitudes of oscillations calculated theoretically $a_{theor}=V/\nu$ and
 from the direct simulation of Eq.~(\ref{NLS})  $a_{num}$ for $\rho_0=1$ and
 different initial values  of the parameter $\vartheta_0$ and for two different trap potentials.}
\begin{ruledtabular}
\begin{tabular}{l|cc|cc}
\hline
$\vartheta_0$  & \multicolumn{2}{c|}{$\nu=0.2$}  & \multicolumn{2}{c}{$\nu=0.1$}
\\
%\cline{2-5}
& $a_{theor}$ & $a_{num}$  & $a_{theor}$ & $a_{num}$
\\
\hline
$1$  &    $8.78$  &  9.17 & 17.55  & 17.99 \\
 1.25 & $8.11$  & 8.33 & 16.22 &  16.59  \\
1.5 & $7.32$ & 7.49 & 14.63 &  14.96  \\
1.75& $6.41$  & 6.56 & 12.82  &  13.09  \\
2 &   $5.40$ & 5.54  & 10.80 & 11.01 \\
2.25& $4.31$ & 4.42 & 8.62 &  8.77  \\
2.5&  $3.15$ & 3.23 & 6.31 &  6.41  \\
2.75& $1.95$ & 1.99 & 3.89 &  3.95  \\
3&    $0.71$ & 0.73 & 1.42 &  1.44  \\
\end{tabular}
\label{tabone}
\end{ruledtabular}
\end{table}

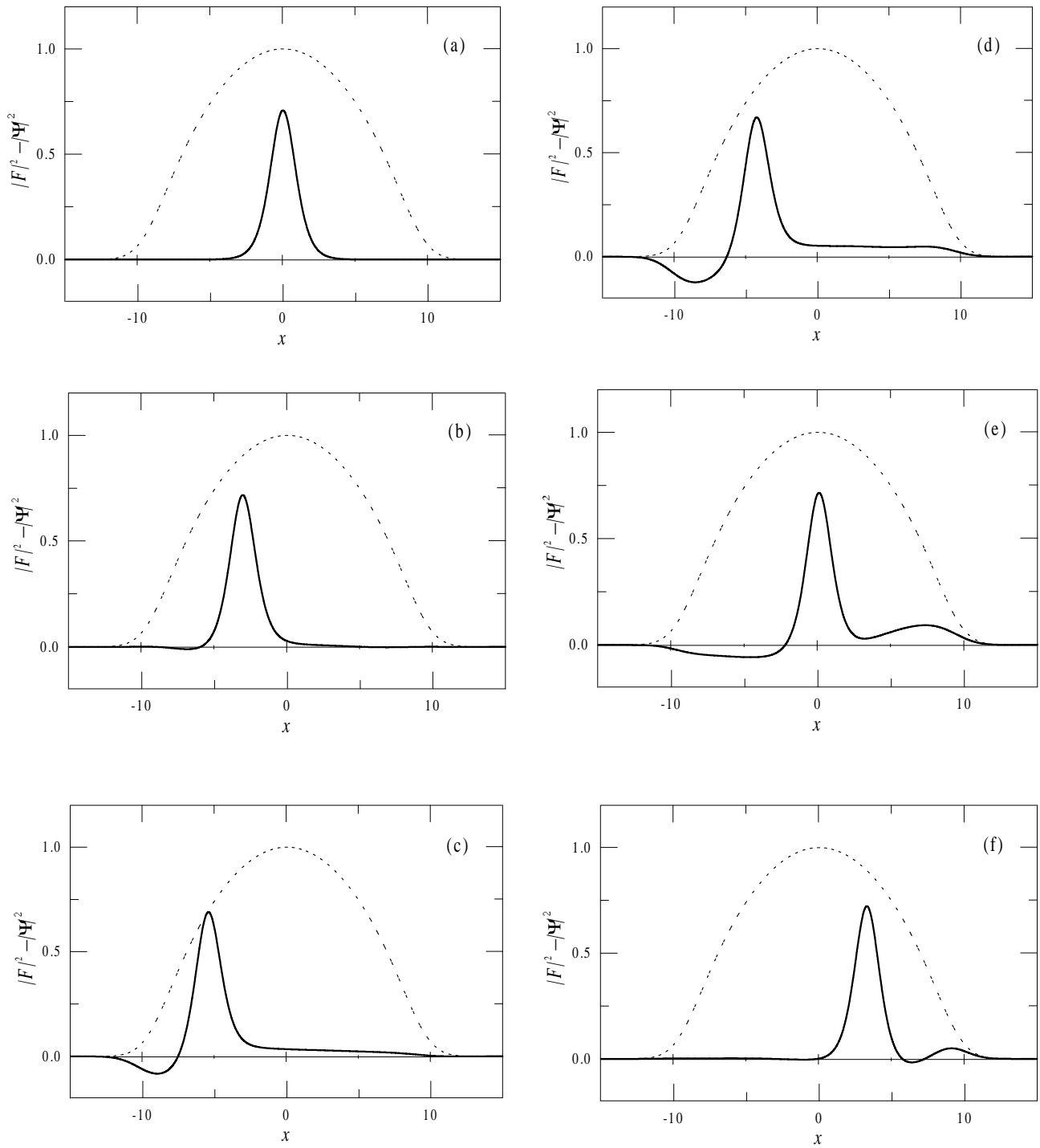
\begin{figure}[h]
%\centerline{\includegraphics[width=4cm,height=2.5cm,clip]{fig2a.eps}\includegraphics[width=4cm,height=2.5cm,clip]{fig2d.eps}}
%\centerline{\includegraphics[width=4cm,height=2.5cm,clip]{fig2b.eps}\includegraphics[width=4cm,height=2.5cm,clip]{fig2e.eps}}
%\centerline{\includegraphics[width=4cm,height=2.5cm,clip]{fig2c.eps}\includegraphics[width=4cm,height=2.5cm,clip]{fig2f.eps}}
\vspace{0.5 cm}
\caption{The reduced shape of the initial distribution computed as $|F(x)|^2-|\Psi(x,t)|^2$ (solid line)
for $\vartheta_0=2$ and $\nu=0.2$ at (a) $t=0$; (b) $t=3$; (c) $t=7.3$; (d) $t=11$; (e) $t=15.2$; (f) $t=18$.
Thin dashed lines show the background $|F(x)|^2$.
}
\label{figtwo}
\end{figure}

In Fig.~\ref{figtwo} we present snapshots of the dark soliton evolution for a
region of parameters when the perturbation theory strictly speaking is not
applicable ($\vartheta_0=2$ and $\rho_0=1$ correspond to $V\approx 1.08$ and thus $V>\nu$).
One can observe rather strong non-adiabatic effects which manifest themselves in
a significant deformation of a soliton shape, namely  in formation of leading and
trailing waves before and behind
the soliton in the ``background'' distribution.

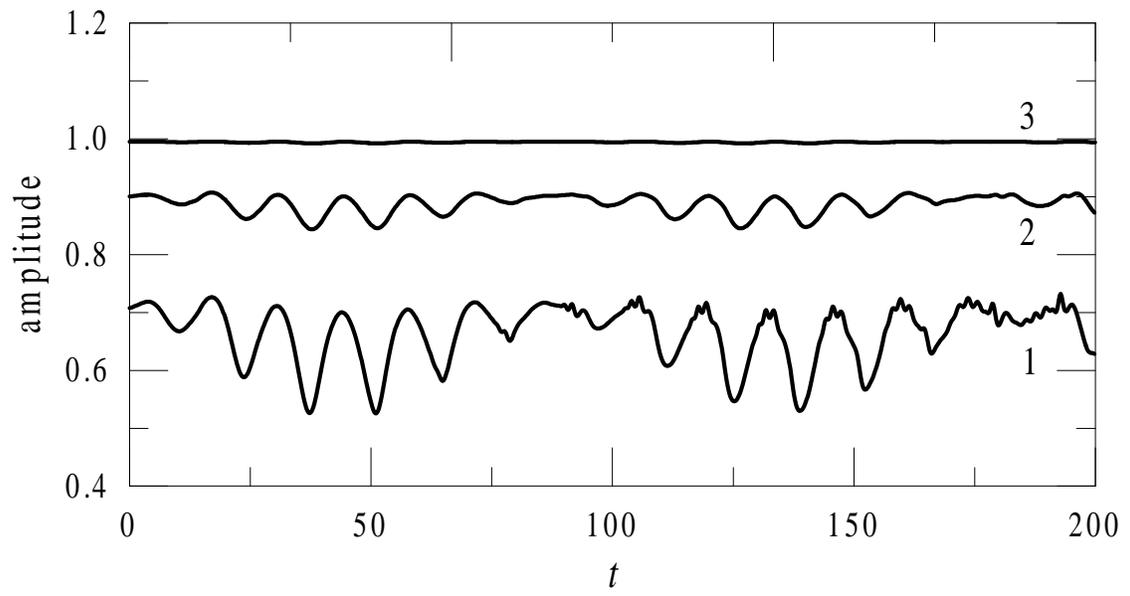
\begin{figure}[h]
\vspace{0.7 cm}
%\centerline{\includegraphics[width=6cm,height=3cm,clip]{fig3_2.eps}}
\vspace{0.8 cm}
\caption{Time dependence of the depth of the dark soliton for $\rho_0=1$, $\nu=0.2$ and
different initial parameters $\vartheta_0=2;2.5;3$ (curves 1,2,3 correspondingly).
}
\label{figthree}
\end{figure}

Some features of the dark soliton dynamics can be understood much better if one
investigates in detail the turning points.
First of all, taking into account that only a black soliton has zero velocity,
it is natural to suppose that at the turning points the soliton becomes black.
The less is the soliton depth, the farer from the center is a turning point,
and hence the greater is the amplitude of the oscillations. This correlation
between variations of
the depth and of the amplitude becomes evident if one compares the results depicted in
Fig.~\ref{figthree} and in Fig.~\ref{figone}. As it has been mentioned in the previous
section, these oscillations are due to mismatch between the the first and higher
harmonics. In all numerical studies carried out we have observed oscillations similar
to ones shown in Fig.~\ref{figone}. They can be explained by the mismatch
$3(\nu-\tilde{\nu})$. We however did not observe pronounced oscillations due
to difference in the second harmonics, i.e. due to $2(\nu-\tilde{\nu})$. We attribute
this behavior to the fact that non-adiabatic effects due to the third harmonics are
effectively much stronger than due to the second one.

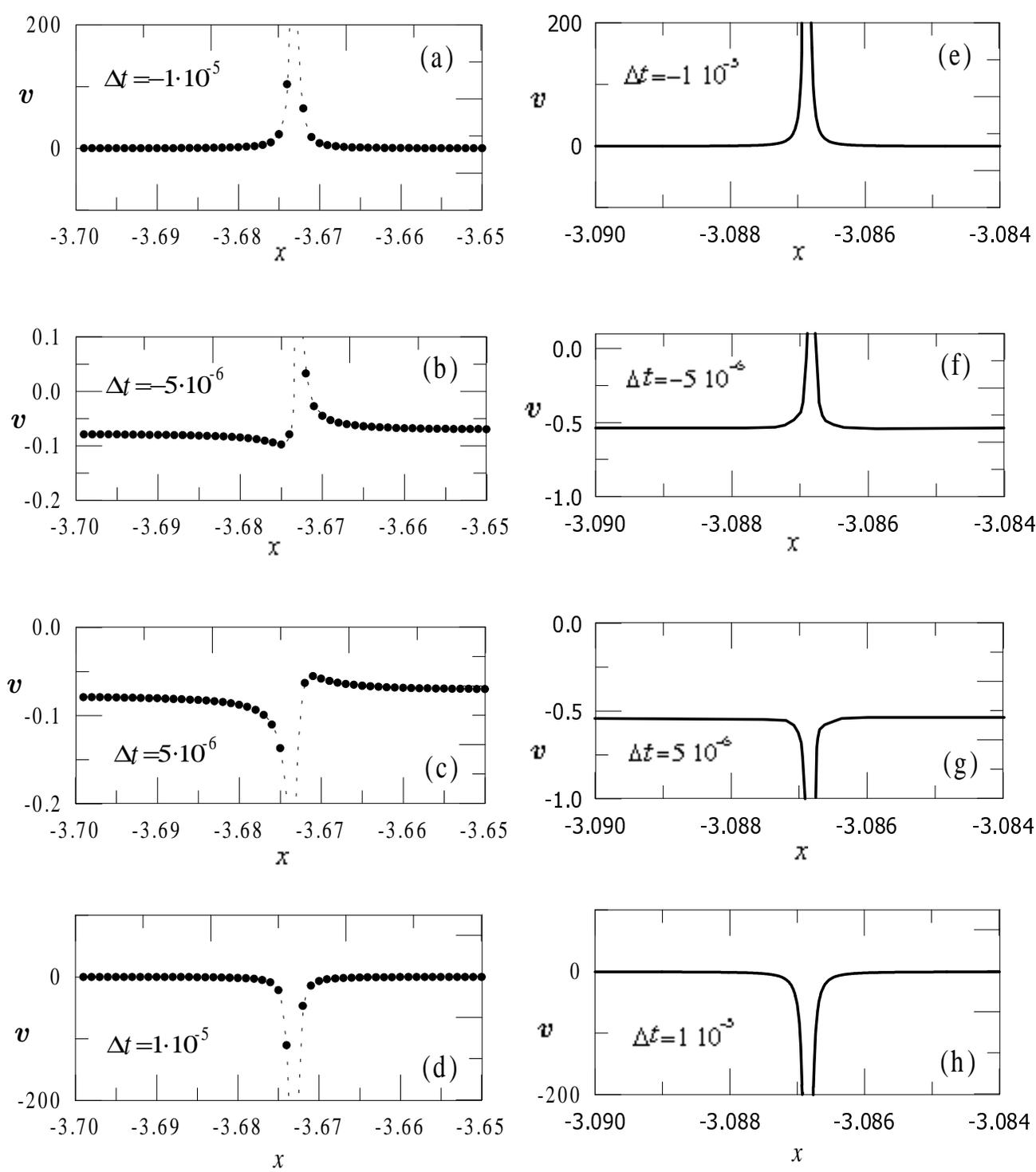
\begin{figure}[h]
%\centerline{\includegraphics[width=4.5cm,height=10cm,clip]{fig4a.eps}
%\includegraphics[width=4.5cm,height=10cm,clip]{fig4b_.eps}}
\vspace{0.5 cm}
\caption{Behavior of the hydrodynamic velocity $v=\partial \varphi_{tot}/\partial x$
in the vicinity of a turning point for
initial parameters $\vartheta_0=2$ and $\nu=0.3$. $\Delta t$ is a time counted from the
turning point.
Left column is an output of direct numerical simulations and right one is formally obtained
form the perturbation theory (see the text).}
\label{figfour}
\end{figure}

Finally, an important effect accompanying changes of the soliton velocity is a change
of the total phase, $\varphi_{tot}=\arg\Phi(x,t)$, of the solution which is especially
significant at the turning points. Our numerical studies of the phase evolution are shown in Fig.~\ref{figfour},
where in the left column we depicted the hydrodynamical velocity $v$ defined as
$v=\partial \varphi_{tot}/\partial x$ (see also the next section).
As one can see, near the turning point hydrodynamic velocity changes
dramatically making a flip (see Figs.~\ref{figfour}b,c),
and after that soliton moves in the opposite direction (Fig.~\ref{figfour}d).
In the right column we show the phase variation during the reflection formally
computed using formulas of the perturbation theory, i.e. Eqs. (\ref{adiab_sol}),
(\ref{ans}), (\ref{expan_phi}), and (\ref{a_Phi}).
Comparing the two sets of graphics
one observes remarkable similarity of the main features (and, what is more important,
the main scales) of the dynamics in the both cases.
Three differences, however, are observable. First, the phase of the direct
numerical solution displays  asymmetry, which is naturally explained by the asymmetric
non-adiabatic  modulation of the soliton shape (c.f. Fig.~\ref{figtwo}).
Second,  spatial locations of the turning points are slightly different. Third,
asymptotic values of the hydrodynamical velocity far from the soliton center differ
as well. These discrepancies are explained  by the fact that the case under consideration
does not satisfy the conditions of the applicability of the adiabatic approximation,
and thus one can expect that the last one can give only right orders of
magnitude of the parameters and indicate correctly the main qualitative features
(what is indeed observed in comparison of the two sets of graphics).

\section{Formation of solitons from large initial excitation}

The problem of evaluation of parameters of dark solitons formed
from a large initial excitation on a constant background is
formally solved by the inverse scattering method~\cite {Zakh}. In
the framework of this method, the NLS equation is associated with
the so-called Zakharov-Shabat linear spectral problem \cite{Zakh},
and soliton parameters are related with the eigenvalues of this
problem calculated for a given initial condensate wave function $\Psi(x,0)$.
If the initial disturbance is large enough, so that the linear
spectral problem possesses many eigenvalues, then a well-known
quasi-classical method can be applied for their calculation. As
was shown in recent paper \cite{Kamch1}, a generalized
Bohr-Sommerfeld quantization rule is very convenient for this aim.

To formulate this rule, it is convenient to introduce a new small
parameter $\eps$, $\varepsilon\ll 1$, into Eq.~(\ref{NLS}) by means of replacements
$x= x'/\eps$, $t= t'/\eps$, $\nu=\nu'\eps$, so that the equation
transforms to
\begin{equation}\label{NLS-eps}
i\eps\frac{\partial\Psi}{\partial t}+\eps^2\frac{\partial^2\Psi}{\partial{x}^2}
-2 |\Psi |^2\Psi =\frac12\nu^2x^2\Psi,
\end{equation}
where we have omitted for simplicity the primes in the new variables.
Then the limit $\eps\ll 1$ corresponds to formation of a large number of
solitons from an initial disturbance with parameters of order of magnitude
$O(1)$ (see \cite{Kamch1}).
In framework of the inverse scattering
transform method the NLS equation, i.e. Eq.~(\ref{NLS-eps}) with the zero
right hand side, is treated
as a compatibility condition of two linear equations for
auxiliary function $\chi$, which we write down in the form
\begin{equation}\label{c2}
  \eps^2\chi_{xx}= \mathcal{A}\chi, \qquad \chi_t=-\frac12 \mathcal{B}
  \chi_x+ \mathcal{B}_x\chi,
\end{equation}
(equivalent to the Zakharov-Shabat problem; see \cite{Alber,KK02}), where
\begin{eqnarray}\label{c3a}
  \mathcal{A}&=&-\left(\la-\frac{i\eps\Psi_x}{2\Psi}\right)^2+|\Psi|^2-
  \eps^2\left(\frac{\Psi_x}{2\Psi}\right)_x,
  \\ \label{c3b}
  \mathcal{B}&=&2\la+\frac{i\eps\Psi_x}{\Psi}.
\end{eqnarray}
The first equation (\ref{c2}) may be considered as a second order scalar
spectral problem with a given ``potential'' $\Psi$ and $\la$ playing the
role of the spectral parameter. When $\Psi(x,t)$ evolves according to
Eq.~(\ref{NLS-eps}) with $\nu=0$, the eigenvalues $\la_n$ of
this problem do not change with time $t$, and each eigenvalue corresponds
to a soliton created from the initial pulse. To investigate the limit $\eps\ll 1$,
let us represent the condensate wave function in the form
\begin{equation}\label{c4}
  \Psi(x,t)=\sqrt{\rho(x,t)}\exp\left(\frac{i}{\eps}\int^x v(x',t)dx'\right),
\end{equation}
where $\rho(x,t)$ has a meaning of the condensate density and $v(x,t)$ is the
hydrodynamic velocity. Indeed, substitution of (\ref{c4}) into
(\ref{NLS-eps}) yields the system
\begin{equation}\label{c5}
  \begin{array}{ll}
  \frac12\rho_t+(\rho v)_x=0,\\
  \frac12v_t+vv_x+\rho_x+\eps^2\left((\rho_{x}^2-2\rho\rho_{xx})/8\rho^2
  \right)_x=-\frac12\nu^2x,
  \end{array}
\end{equation}
which for $\eps\to 0$ takes the form of hydrodynamic equations.
For smooth enough functions $\rho(x,t)$ and $v(x,t)$, when
\begin{equation}
\label{cond_WKB}
|\eps\rho_x/\rho|\ll \rho,\quad \mbox{and}\quad  |\eps v_x|\ll\rho,
\end{equation}
what corresponds to neglecting the space derivatives in $\mathcal{A}$,
spectral problem (\ref{c2}) transforms into
\begin{equation}\label{c6}
  \eps^2\chi_{xx}=\left[-\left(\la+v/2\right)^2+\rho\right]\chi.
\end{equation}
It is to be mentioned here that conditions (\ref{cond_WKB}) are nothing but
the conditions of applicability of the well-known hydrodynamical
approach when due to a relatively high density the two-body interactions
are strong enough and one can neglect the ``quantum pressure''
(see, e.g. \cite{Pit}).

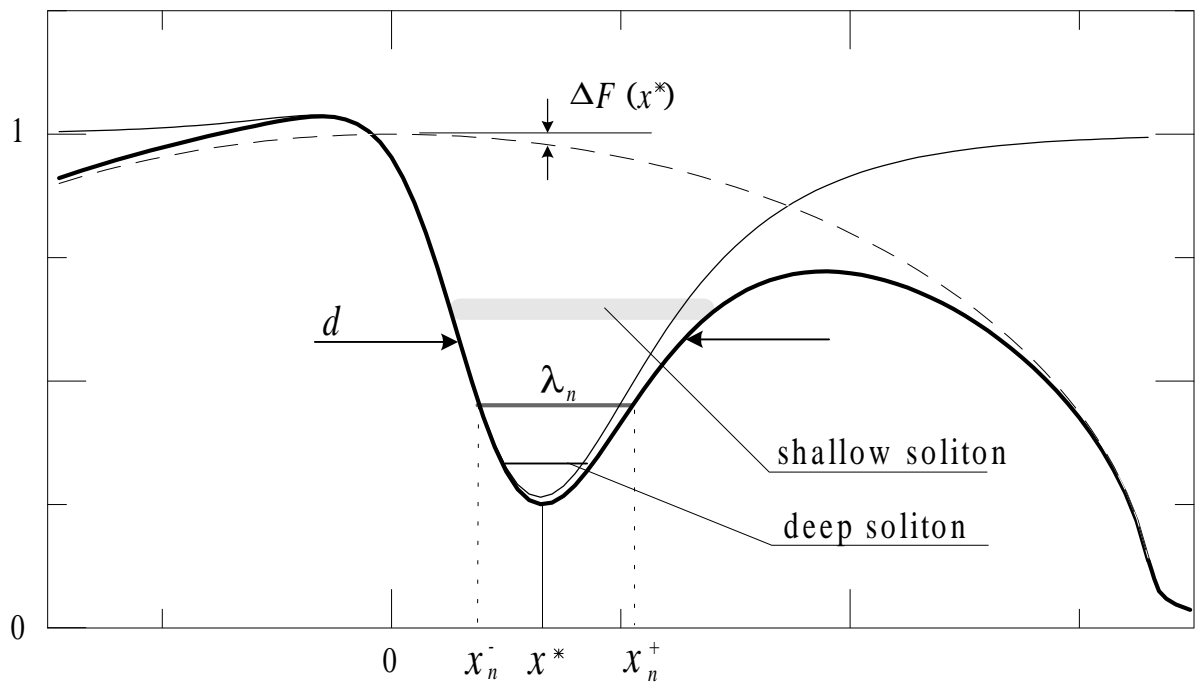
\begin{figure}[h]
%\centerline{\includegraphics[width=8cm,height=5cm,clip]{fig_n.eps}}
\caption{Schematic plot of the Riemann invariant $\la^+$ given by Eq.(\ref{riemann})
(thick solid line); thin solid line shows the Riemann invariant for disturbance
with respect to uniform background; dashed line shows background without disturbance.
$x^*$ denotes the position of localized disturbance and
$\Delta F(x^*)=1-F(x^*)$
is the change of the condensate density because of condensate non-uniformity.
$d$ is characteristic scale of the initial disturbance.
The horizontal lines of different width   indicate positions of eigenvalues
$\la_n$ and the width of each line characterizes the lifetime of  the corresponding solitons
(the thicker is a line, the smaller is the lifetime).
}
\label{fig_n}
\end{figure}

Equation (\ref{c6}) has a formal analogy with a stationary Schr\"odinger
equation for a quantum-mechanical motion of a particle in the
``energy-dependent'' potential, i.e. the potential depending on the
spectral parameter $\la$. According the mentioned above independence
of the eigenvalues $\la_n$ on time, they can be calculated with the
use of the initial distributions $\rho(x,0)$ and $v(x,0)$. Since we
are interested in such initial data which give rise to creation of
large number of solitons, this means that functions $\rho(x,0)$ and $v(x,0)$
correspond to the problem (\ref{c6}) with a large number of eigenvalues.
Then the quasi-classical approach can be used for their calculation
\cite{Kamch1}. To clarify this method, we have shown schematically
in Fig.~\ref{fig_n} a plot of the ``Riemann invariant''
\begin{equation}\label{riemann}
  \la^+=-v(x,0)/2+\sqrt{\rho(x,0)}
\end{equation}
which plays a role of the ``quantum-mechanical potential'' for the
problem (\ref{c6}). (The second Riemann invariant $\la^-=-v(x,0)/2-\sqrt{\rho(x,0)}$
can be considered in the same way.) The Riemann invariant for the same
disturbance but with respect to uniform background $F(x)=1$ is shown
for comparison by thin solid line. We see that in both cases there is a
``potential well'', but for the nonuniform background case the eigenvalues
acquire imaginary part (``decay width'') due to tunnelling effect.
This means that dark solitons in confined condensate have a finite life-time
$\tau_n$ determined by the imaginary part of the eigenvalue $\la_n$,
what correlates with the above mentioned fact that in this case the
soliton is not a ``well-defined'' object. Nevertheless, it makes
sense to speak about solitons in a confined condensate, if their
life-times $\tau_n$ are much greater than the period $\simeq2\pi/\nu$
of their oscillations discussed in the preceding Section. It is clear
that``shallow'' solitons with small $\tau_n$ do not survive in the
confined condensate, so that $\la_n$ with values close to the top
of the ``potential barrier'' do not correspond to any real solitons.
On the contrary, in the case of the uniform background discussed in
\cite{Kamch1} all eigenvalues correspond to real solitons appearing
eventually from the initial pulse. In multi-soliton problem there is also
one more scale of time equal to time of formation of solitons from
the initial pulse. For deep enough solitons it can be estimated by
the order of magnitude as time necessary for solitons with velocity
$|V_n|=2|\la_n|$ (see below) to pass the distance equal to
the width $d$ of the initial problem. For the problems under
consideration, this time $\sim d/2|\la_n|$ must be much less than
the period $2\pi/\nu$. Thus, we are interested in the eigenvalues
$\la_n$ which satisfy inequalities
\begin{equation}\label{ineq}
  \tau_n\gg\frac{2\pi}{\nu}\gg\frac{d}{2|\la_n|}.
\end{equation}
It is clear that there are no these reservations in the case of
uniform background \cite{Kamch1} where formally $\tau_n=\infty$ and one
can wait long enough to observe formation of soliton with any value of
$\la_n$.

To calculate the real parts of the eigenvalues $\la_n$ corresponding to
deep solitons, one can use the generalized Bohr-Sommerfeld quantization
rule
\begin{eqnarray}\label{eq14}
&& \frac1{\eps}\int^{x^+_n}_{x^-_n} \sqrt{\left(\la_n+\frac12v(x,0)\right)^2
-\rho(x,0)} \,dx=\pi\left(n+\frac12\right),\nonumber \\
&&\quad n=0,1,2,\ldots,M,
\end{eqnarray}
with given initial distributions $\rho(x,0)$ and $v(x,0)$.
We suppose here that the integrand has only one maximum and $x^+_n$
and $x^-_n$ are the points where the integrand function vanishes: they depend on
$\la_n$ and are chosen so that the relationship (\ref{eq14}) is satisfied
(see Fig.~\ref{fig_n}).
Analytical form of each emerging soliton in an asymptotic region
where it is well separated from other solitons (i.e. in the limit $t\to\infty$)
is expressed in terms of $\la_n$ as follows:
\begin{equation}\label{c7}
    \rho_s^{(n)}(x,t)=\rho_0-\frac{\rho_0-\la^2_n}{\cosh^2\left[
    \sqrt{\rho_0-\la^2_n}\,(x-2\la_nt)/\eps\right]},
\end{equation}
\begin{equation}\label{c8}
  v_s^{(n)}(x,t)=\la_n\left(\rho_0/\rho_s^{(n)}-1\right).
\end{equation}
As it is clear, formulas (\ref{c7}), (\ref{c8}) and (\ref{adiab_sol}) with
$\vartheta=\vartheta_n$, $\eta_0=\eta_n$, and $V=V_n$, where the parameters
are connected by the relation $\lambda_n=\sqrt{\rho_0}\cos(\vartheta_n/2)$,
represent the same one-soliton solution.
Thus, the last formula allows one to find initial values of $\vartheta_n$ for
solitons emerging from the dark excitation of the
condensate with given initial distributions of $\rho(x,0)$ and
$v(x,0)$ against a constant uniform background.

If the background is not uniform but
changes in space in the intervals of integration in (\ref{eq14}),
then we can apply the
same method with $\rho_0$ replaced by the value of the background
density $F^2(x^*)$ at the place $x^*$ of the localized initial excitation
(see Fig.~\ref{fig_n}).

We have used this approach for finding soliton parameters for different
types of initial excitations: (i) excitations of the density $\rho(x)$ \cite{andrews},
(ii) excitation of the hydrodynamic velocity $v(x)$ (``phase imprinting method'')
\cite{Burger1,Biao} and (iii) collision of condensates \cite{Scott}.

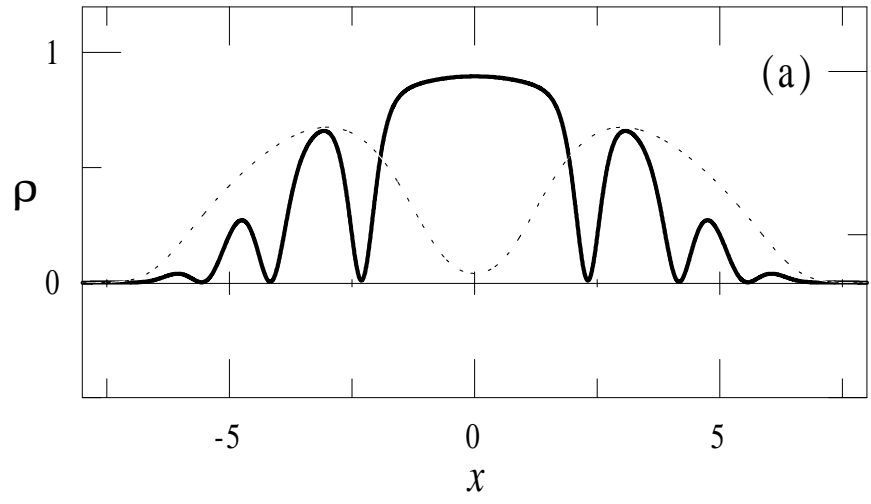
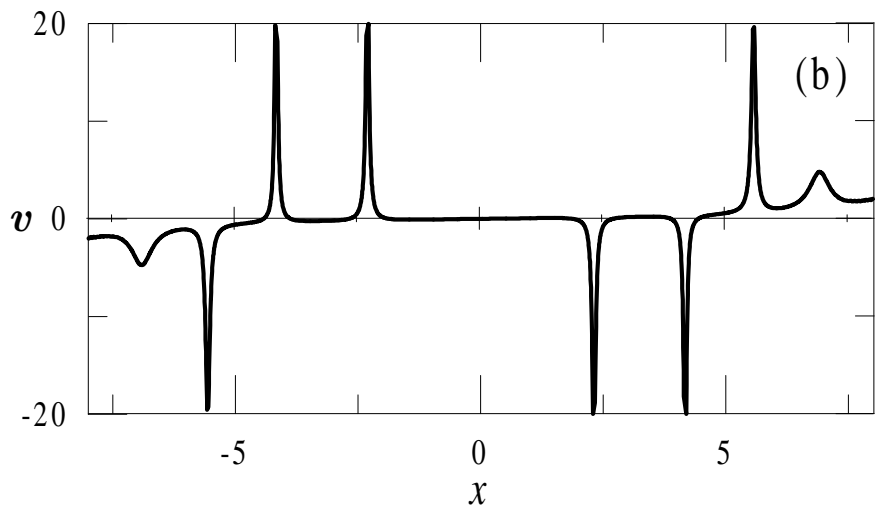
\begin{figure}[h]
%\centerline{\includegraphics[width=6cm,height=6cm,clip]{fig5.eps}}
\caption{Space distribution of the density of a BEC (a) and of its hydrodynamic velocity (b)
in the harmonic trapped potential with $\nu=0.3$
at time $t=4.24$ with initial excitation taken in the form of pulse (\ref{eq17})
with  $\rho_0=1$, $\alpha=0.8$ and $\varepsilon=0.3$.}
\label{figfive}
\end{figure}

(i) In the first case of the density disturbance the initial data were
taken in the form
\begin{equation}\label{eq17}
  \rho(x,0)=\left( 1- \frac{\alpha}{\cosh(x)}\right)^2,\quad v(x,0)=0,
\end{equation}
where the parameter $\alpha$ measures the strength of the disturbance.
We have chosen the following values of the parameters:
$\alpha=0.8$, $\nu=0.2$, $\eps=0.3$. The values of
$\la_{n}$ for the three deepest solitons calculated with the use the
Bohr-Sommerfeld rule (\ref{eq14}) are shown in Table~\ref{tabtwo} together with
the corresponding values of $\vartheta_{n}$ and amplitudes $a_{theor}^{(n)}$
calculated for these soliton from (\ref{sol}) and found from numerical
solution of (\ref{NLS}) with the initial data (\ref{eq17}). The
discrepancy is less than 10\% and is caused, apparently, by the fact
that in our case the created solitons did not reach yet the asymptotic values
of their velocities $V_n=2\la_n$.
Besides that, numerical calculations show that the ``initial coordinates''
of solitons created from the initial pulse cannot be identified exactly
with $X(0)=0$. This is the reason why solitons created from one initial pulse do
not reverse simultaneously their directions of motion even during the first
period of oscillations in the confined condensate.
This phenomenon is
illustrated in Fig.~\ref{figfive}, where the density $\rho(x,t)$ and hydrodynamic
velocity $v(x,t)$ of the condensate are shown as functions of $x$ at
the moments when two solitons in each train have already reversed
the direction of their propagation and are moving to the center,
while the other two solitons are still moving from the center.
This process of ``solitons reflection from the potential well''
takes about 20\% of the whole period $2\pi/\nu$ of their oscillations.

\begin{table}
\caption{ Parameters of solitons created from initial intensity disturbance}
\begin{ruledtabular}
\begin{tabular}{c|cccc}
\hline
 \ \  $n$  \ \ & \ \ $\lambda^{(n)}$  \ \ &  \ \ $\vartheta^{(n)}$  \ \ &
  \ \ $a^{(n)}_{theor}$  \ \ &  \ \ $a^{(n)}_{num}$ \ \ \\ \hline
  0 & 0.41 & 2.29 & 2.74 & 2.48\\
  1 & 0.65 & 1.72 & 4.35 & 4.23\\
  2 & 0.80 & 1.28 & 5.42 & 5.54\\
\end{tabular}
\label{tabtwo}
\end{ruledtabular}
\end{table}

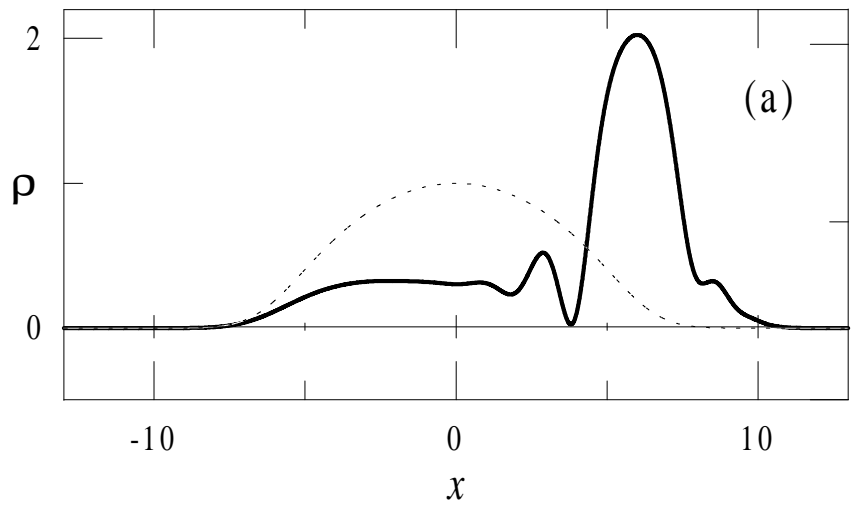
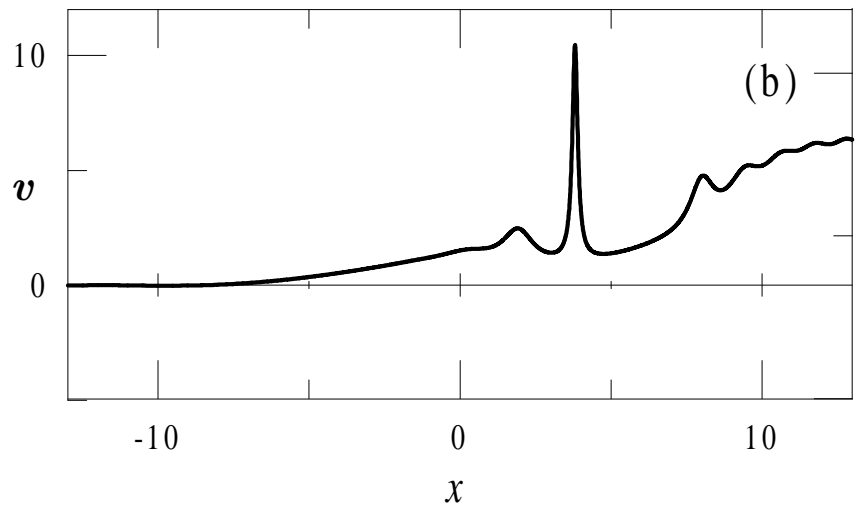
\begin{figure}[h]
%\centerline{\includegraphics[width=6cm,height=6cm,clip]{fig6.eps}}
\caption{Space distribution of density of BEC (a) and its hydrodynamic velocity (b)
in the harmonic trapped potential with $\nu=0.3$ at time $t=1$ without (dotted line)
and with (solid line) initial excitation that is taken as a phase step
with  $\rho_0=1$, $\alpha=4$, $\kappa=0.4$ and $\varepsilon=1$.
}
\label{figsix}
\end{figure}

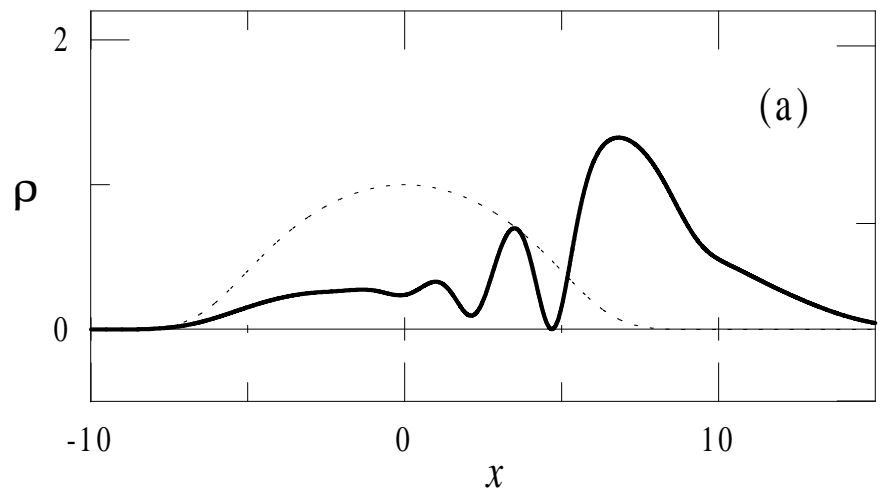
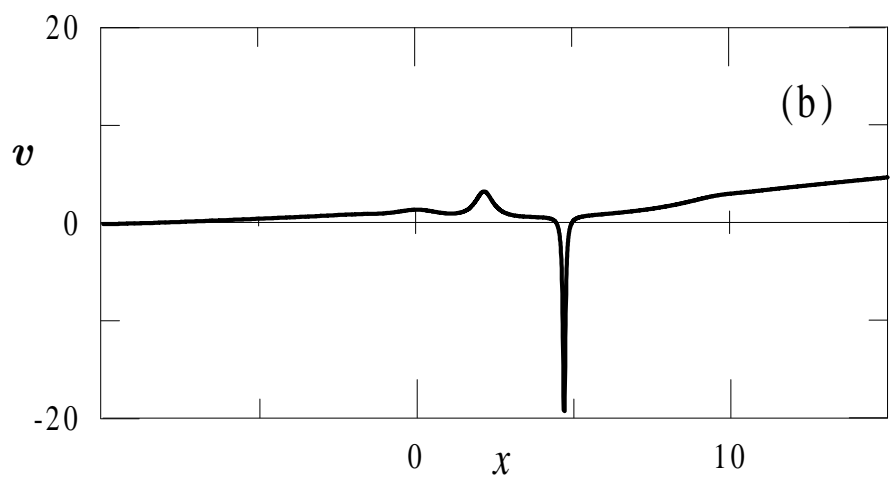
\begin{figure}[h]
%\centerline{\includegraphics[width=6cm,height=6cm,clip]{fig7.eps}}
\caption{The same as on Fig.~\ref{figsix} at $t=1.5$.
}
\label{figseven}
\end{figure}

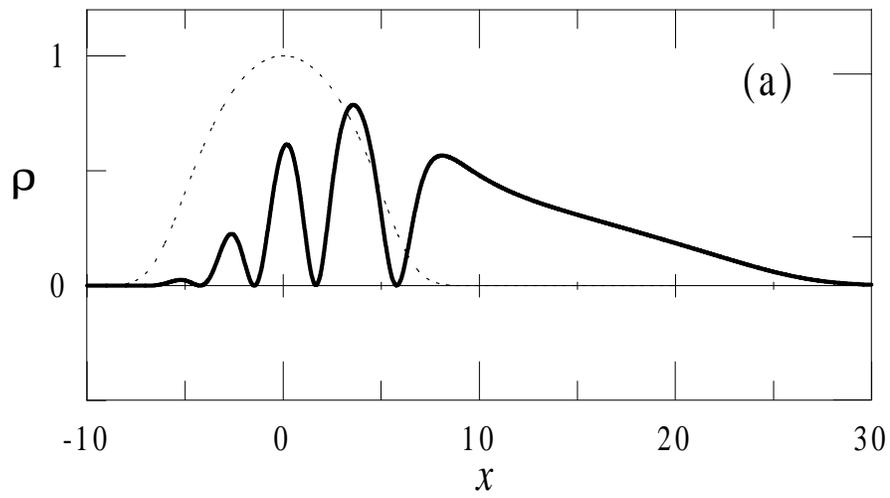
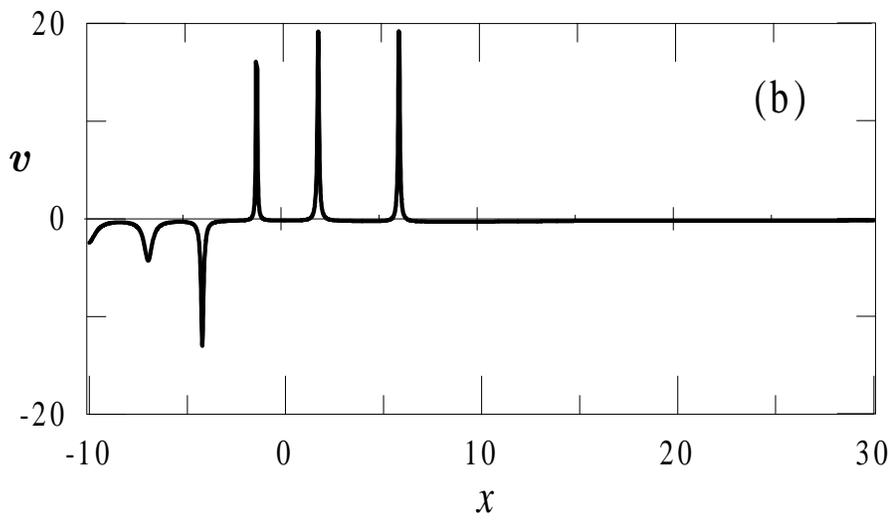
\begin{figure}[h]
%\centerline{\includegraphics[width=6cm,height=6cm,clip]{fig8.eps}}
\caption{The same as on Fig.~\ref{figsix} at $t=4$.
}
\label{figseight}
\end{figure}

(ii) For illustration of the process of solitons formation by the phase
imprinting method we have chosen the initial conditions in the form
\begin{equation}\label{eq18}
  \rho(x,0)=F^2(x),\qquad v(x,0)=\frac{\alpha}{\cosh^2(\kappa x)}.
\end{equation}
Again the solitons parameters can be calculated with the use of the
Bohr-Sommerfeld quantization rule (\ref{eq14}) and their values
correspond well to numerical simulation. In particular, the
Bohr-Sommerfeld rule gives correct number of solitons and signs of their
initial velocities.
As one can see in Fig.~\ref{figsix}, at first the deepest soliton moves to the right
but the hydrodynamic velocity distribution corresponds to its motion
to the left. Only when the local density minimum touches the $x$ axis,
the velocity distribution makes a flip and after that it corresponds to
the predicted direction of the soliton propagation (see Fig.~\ref{figseven}).
All solitons  predicted by the Bohr-Sommerfeld quantization rule can be observed
during some
time interval after their formation (see Figs.~\ref{figseven},\ref{figseight}).
However, in the case of initial data (\ref{eq18}) we observe also strong
non-soliton contribution into excitation of the condensate (see Fig.~\ref{figsix})
which leads to much more complicated picture of its evolution. One may
say that in this case solitons move along background varying with
time and the influence of this time dependence is not small on the
contrary to the previous case of the density disturbance. As a result,
the motion of solitons cannot be described as (almost) harmonic
oscillations along constant nonuniform background. But even in this case
quasi-classical method of finding the solitons parameters yields
quite accurate description of solitons motion for times less than
$2\pi/\nu$.

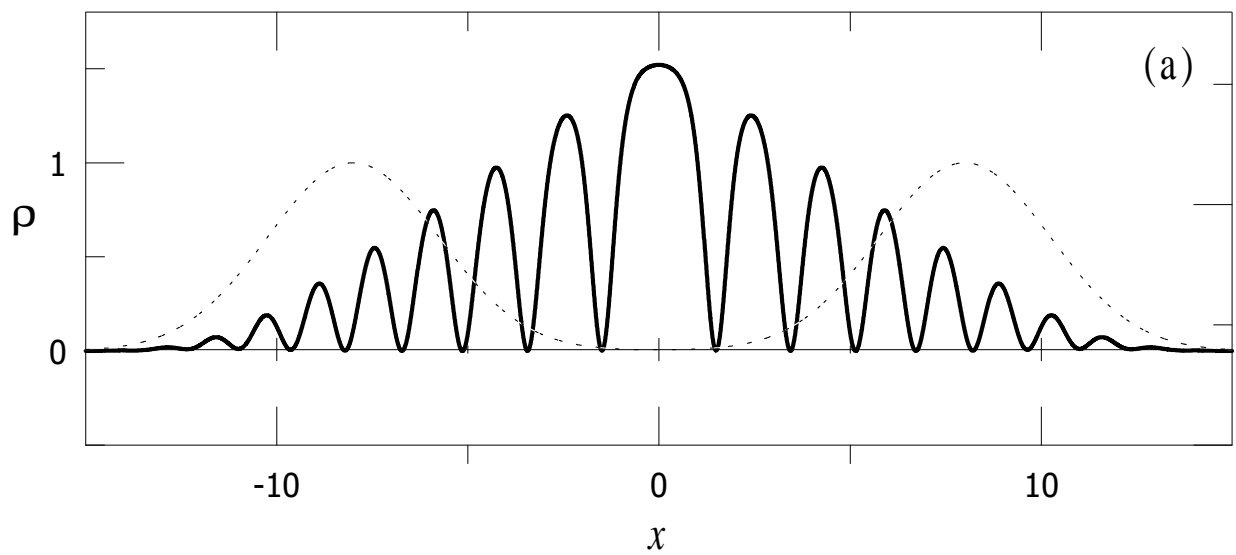
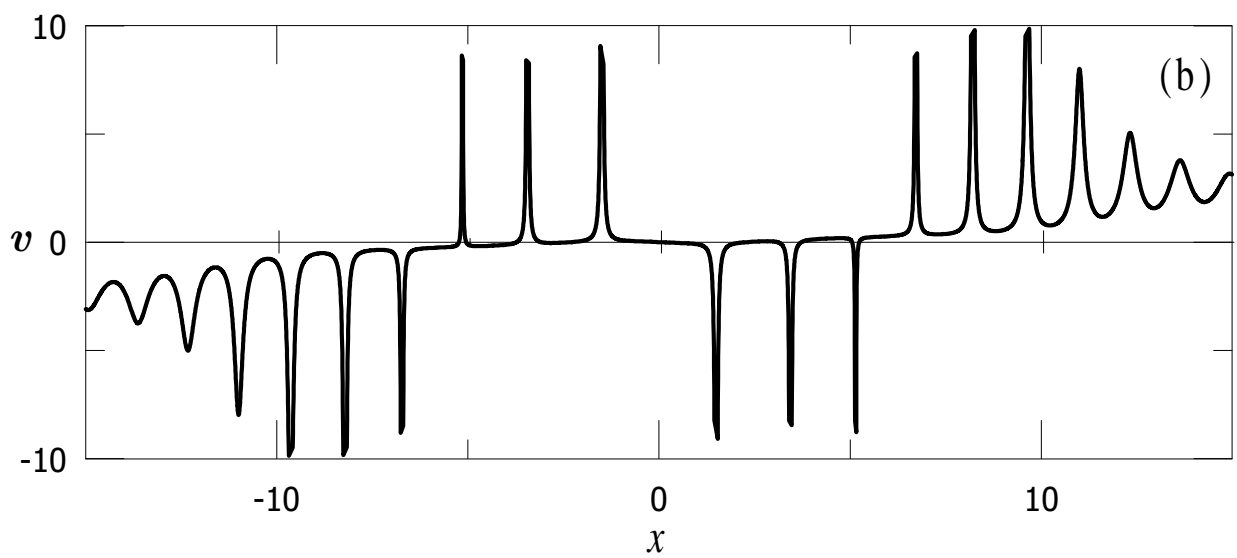
\begin{figure}[h]
%\centerline{\includegraphics[width=6cm,height=6cm,clip]{fig9.eps}}
\caption{Formation of dark solitons (solid line)
during the collision of two initially separated condensates
(dotted line) corresponding to the initial data (\ref{coll1}) with parameters
$\kappa=0.5$, $\xi=8$ and $\varepsilon=0.5$
in harmonic trap potential $\nu=0.2$ at $t=4$.
a) density of BEC; b) hydrodynamic velocity.
}
\label{fignine}
\end{figure}

(iii) Formation of dark solitons can be observed also during collision of two separated
condensates which move under influence of the trap potential \cite{Scott}.
Taking the initial conditions in the form
\begin{eqnarray}
\rho(x,0)&=&\exp\left[-
\kappa(x-\xi)^2\right]+\exp\left[-{\kappa(x+\xi)^2}\right],\nonumber \\
\label{coll1}
v(x,0)&=&0
\end{eqnarray}
we have solved  Eq.~(\ref{NLS}) numerically. It was found that these two condensates
oscillate in the trap potential and interact with each other in quite complicated way when
they overlap with each other. Since the initial data (\ref{coll1}) lead to a number
of eigenvalues $\lambda^{(n)}$, one may expect that during the collision of two
condensates the corresponding number of dark solitons must be observed.
This is indeed the case as one can see in Fig.~\ref{fignine}, where the density and the
hydrodynamic velocity distributions are shown at the moment of maximal
overlap of two condensates whose initial density distributions are
indicated by dotted lines. The number of solitons matches very well with that
~predicted by the Bohr-Sommerfeld quantization rule, but their motion cannot
be presented as propagation with slowly changing parameters along
constant nonuniform background.

Thus, quasi-classical approach provides simple and effective method
of calculation of the parameters of solitons arising from large enough
initial disturbance. This method can be used for estimation of these
parameters in the today experiments with BEC solitons.

\section{Discussion and conclusion}

In the present paper we have investigated the evolution of
localized excitations in a quasi-1D BEC with a positive scattering length
confined by a harmonic trap potential. It has been shown that in
the case of a single dark soliton the evolution can be considered
as a newtonian one only at very low velocities and, hence, big depths of
the soliton and at a large enough effective longitudinal size of the
condensate. Then the dynamics becomes near-integrable and the
perturbation theory for dark solitons can be used. Non-adiabatic
effects become essential when one considers long-time dynamics,
i.e. dynamics when soliton makes several oscillations. These
effects are: increase of the frequency of  soliton oscillations;
increase of the amplitude of oscillations, which, besides that,
changes periodically with a frequency much less than the frequency
of the oscillations themselves; change of a soliton shape during
the evolution.

Main effects observed in non-newtonian dynamics of a soliton are
described qualitatively by the perturbation theory for dark
solitons. In particular, the theory  allows one to justify the
choice of the shape of the background which supports
many-cycle dynamics of a soliton without substantial change of
its shape.
%Also the theory provides us with an explicit form
%of a dark soliton in the leading order with respect to the small
%parameter $\nu$.
%
The present study, however, leaves open a question about
non-adiabatic deformation of a soliton shape. This requires careful
study of the first order perturbation theory which can be
made, say, with the use the Green function approach (see \cite{KV94}) for the
respective linearized problem.

The existence of an inhomogeneous background becomes especially
important when initially multi-soliton pulses are under
consideration. Then, compared to the integrable case with constant
background, new temporal scales appear in the problem. They
are associated with the harmonic oscillator frequency and
finite life-time of solitons. The life-time decreases with the
soliton amplitude, which leads to rather rapid disappearance of
shallow dark solitons. Generally speaking, a soliton with a small
amplitude can even loose its meaning at all when it is considered against
a nonuniform background. Indeed, returning to the condition of
smallness of the soliton width $\sim 1/\eta$ compared with the linear
oscillator length $\sim 1/\nu$, one must require $\eta\gg\nu$ for a
soliton to be meaningful. Recalling now that $a_{theor}\leq
\sqrt{\rho_0}/\nu$ where $a_{theor}$ is an amplitude of soliton
oscillations, what follows from the Ehrenfest theorem, one concludes
that it make sense to speak about the oscillations of a soliton in
the case when $a_{theor}<1/\nu$, i. e. when  $\rho_0 \sim 1$.   If
$\rho_0$ is large enough, then the small amplitude soliton can
reach a region of an exponentially decaying tail of the condensate
distribution already during the first half-a-period of
oscillations. In that region the dynamics is essentially linear
and thus the pulse will disappear due to dispersion effects as it happens
with linear wave-packets. In order to estimate amplitudes of the
background at which such behavior is observable, we take into
account that the effective nonlinearity, determined from
(\ref{pert_NSE}), can be estimated as
$\rho_0\exp(-\nu^2a_{theor}^2)$ and thus it become of order of
$10^{-2}$ already at $\rho_0\approx 6.5$.

%In the case of a multi-soliton initial pulse the Ehrenfest theorem
%becomes much less informative than in the case of a single
%soliton. This is due to the fact that now each of solitons
%emerging from a large initial pulse has different initial
%coordinate (which can be formally defined and which depends on a
%particular shape of the pulse). As a result different solitons in
%a train are reflected by the potential wall not only at different
%turning points but also at different moments of time.

In the context of the above findings a natural question arises
about detecting soliton parameters. In connection with this question
it is relevant to mention that motion of the  soliton is
accompanied with the hydrodynamic flow with velocity $v(x,t)$ and
the corresponding matter current, which density in the dimensionless units
is $J(x,t)=\rho(x,t)v(x,t)$. Dependence of these two quantities
on time is periodic. Their dependence on the spatial coordinate
is also non-monotonic.
%As it is clear,  bosons, considered as
%quasi-particles, has maximal velocity in the point where soliton
%depth reaches its maximum. However, because of the decrease of the
%condensate density in that region the maximum current is reached
%for an intermediate soliton depth. Namely, it is a simple algebra
%to verify that the largest current is achieved by a soliton having
%the velocity $V=\sqrt{\rho_0/3}$ (and respectively the depth
%$\eta=2\sqrt{2\rho_0/3}$).
Thus, by detection of the velocity and current distribution allows
one to make an estimate of the dark soliton parameters.

\subsection*{Acknowledgements}

A.M.K. is grateful to the staff of Centro de F\'{\i}sica da Mat\'eria
Condensada, Universidade de Lisboa, for kind hospitality.
The work of V.A.B. has been supported by the FCT fellowship SFRH/BPD/5632/2001.
The work of A.M.K. in Lisbon has been supported by the Senior NATO Fellowship.
A.M.K. thanks also RFBR (Grant 01--01--00696) for partial support.
V.V.K. acknowledges support from the Programme ``Human Potential-Research Training Networks",
contract No. HPRN-CT-2000-00158.

%\newpage
\appendix
\section{}

For the sake of completeness here we reproduce the dynamical equation for
$x_0(t)$ which is derived in \cite{KV94} (notice that the sign between the
second term in the right hand side, is corrected):
\begin{eqnarray}
\frac {{\rm d}x_0}{{\rm d}t}&=&\int_{-\infty}^{\infty} {\rm d}\Theta\left\{
\frac{1}{\eta_0\eta}\widehat{R}^\prime_+(\Theta,t)
\right.\nonumber \\
&+& \left. \frac{V}{\eta_0^2\eta}\left(
\frac{\Theta/2}{\cosh^2(\frac{\Theta}{2})} +\tanh\frac{\Theta}{2}\right)
\widehat{R}^{\prime\prime}_- (\Theta,t)\right. \nonumber \\
&-&\left.  \frac{4\rho_0\eta}{\eta_0^3}\int_0^t {\rm d}t^\prime
\frac{\widehat{R}^{\prime\prime}_+ (\Theta,t^\prime)}{\sinh\Theta}
\left[\left(1-\frac{V^2}{\rho_0}\right)\tanh\frac{\Theta}{2}
\right. \right. \nonumber \\
&+&
\left.\left.
\frac{V^2}{4\rho_0}\Theta\left(1-\frac32\frac{1}{\cosh^2(\frac{\Theta}{2})}\right)\right]
\right\}
\label{a1}
\end{eqnarray}
where notations
\begin{eqnarray}\label{a4}
\widehat{R}^\prime_\pm(\Theta,t)&=&\frac12{\rm Re} \left[\bar{\eps}\widetilde{R}(\Theta,t)\pm
\eps\bar{\widetilde{R}}(-\Theta,t)\right],
\\
\label{a5}
\widehat{R}^{\prime\prime}_\pm(\Theta,t)&=&\frac12{\rm Im}
\left[\bar{\eps}\widetilde{R}(\Theta,t)\mp
\eps\bar{\widetilde{R}}(-\Theta,t)\right].
\end{eqnarray}
Here $\widetilde{R}$ is determined by Eq.(\ref{R1}).

\end{document}